\documentclass[twocolumn,aps,prb,superscriptaddress,floatfix]{revtex4}
\usepackage[T1]{fontenc}
\usepackage[utf8]{inputenc}
\usepackage[english]{babel}
\usepackage{comment}
\usepackage{color}
\usepackage{float}
\usepackage{braket}
\usepackage{amsmath}
\usepackage{amstext}
\usepackage{amssymb}
\usepackage{graphicx}
\usepackage{bbold}
\usepackage[unicode=true,pdfusetitle,
bookmarks=true,bookmarksnumbered=false,bookmarksopen=false,
breaklinks=false,pdfborder={0 0 1},backref=false,colorlinks=false]
{hyperref}

\hypersetup{
	colorlinks,linkcolor=blue,citecolor=blue,urlcolor=blue}
\usepackage{soul}

\setul{0.5ex}{0.3ex}
\definecolor{Blue}{rgb}{0,0.0,1}
\setulcolor{Blue}
\usepackage{babel}

\begin{document} 

\author{Marcio Costa}\email{mjtcosta@id.uff.br}
\affiliation{Instituto de F\'\i sica, Universidade Federal Fluminense, 24210-346 Niter\'oi RJ, Brazil} 
 
\author{Bruno Focassio}
\affiliation{Federal University of ABC (UFABC), 09210-580 Santo André , São Paulo, Brazil}
\affiliation{Ilum School of Science,CNPEM, 13083-970 Campinas, São Paulo, Brazil}

\author{Luis M. Canonico}
\affiliation{Catalan Institute of Nanoscience and Nanotechnology (ICN2), CSIC and BIST, Campus UAB, Bellaterra, 08193 Barcelona, Spain}

\author{Tarik P. Cysne}
\affiliation{Instituto de F\'\i sica, Universidade Federal Fluminense, 24210-346 Niter\'oi RJ, Brazil}

\author{Gabriel R. Schleder}
\affiliation{John A. Paulson School of Engineering and Applied Sciences, Harvard University, Cambridge, Massachusetts 02138, USA}

\author{R. B. Muniz}
\affiliation{Instituto de F\'\i sica, Universidade Federal Fluminense, 24210-346 Niter\'oi RJ, Brazil}

\author{Adalberto Fazzio}
\affiliation{Federal University of ABC (UFABC), 09210-580 Santo André , São Paulo, Brazil}
\affiliation{Ilum School of Science,CNPEM, 13083-970 Campinas, São Paulo, Brazil}
\author{Tatiana G. Rappoport}
\affiliation{Instituto de Telecomunicações, Instituto Superior Tecnico, University of Lisbon, Avenida Rovisco Pais 1, Lisboa, 1049001 Portugal}	
\affiliation{Instituto de F\'\i sica, Universidade Federal do Rio de Janeiro, C.P. 68528, 21941-972 Rio de Janeiro RJ, Brazil}

\title{Connecting Higher-Order Topology with the Orbital Hall Effect \\ in Monolayers of Transition Metal Dichalcogenides}

\begin{abstract}
Monolayers of transition metal dichalcogenides (TMDs) in the 2H structural phase have been recently classified as higher-order topological insulators (HOTI), protected by $C_3$ rotation symmetry. 
In addition, theoretical calculations show an orbital Hall plateau in the insulating gap of TMDs, characterized by an orbital Chern number.
We explore the correlation between these two phenomena in TMD monolayers in two structural phases: the noncentrosymmetric 2H and the centrosymmetric 1T.  
Using density functional theory, we confirm the characteristics of 2H-TMDs and reveal that 1T-TMDs are identified by a $\mathbb{Z}_4$ topological invariant.  
As a result, when cut along appropriate directions, they host conducting edge-states, which cross their bulk energy-band gaps and can transport orbital angular momentum. 
Our linear response calculations thus indicate that the HOTI phase is accompanied by an orbital Hall effect.
Using general symmetry arguments, we establish a connection between the two phenomena with potential implications for orbitronics and spin-orbitronics.
\end{abstract}

\maketitle
{\it Introduction:} 
The orbital Hall effect (OHE) refers to the transverse flow of orbital angular momentum (OAM) in response to a longitudinally applied electric field. It resembles the spin Hall effect (SHE) \cite{Go-Review,OHEBernevig,OrbitalTexture, OHEmetals}, but unlike the latter, it does not require spin-orbit coupling (SOC). 
Characteristics of the OHE and the physical mechanisms underlying it are currently under investigation ~\cite{Oppener-Salemi-1,Oppener-Salemi-2,Oppener-Salemi-3,Sayantika-OHE-Model-3D,Negative-intrinsic-OHE,Orientational-dependence-OHE}. For instance, signatures of the OHE in 3D metallic systems were recently observed \cite{Exp-OHE-1,orbital-torque-magnetic-bilayers-EXP}, paving the way for possible orbitronic applications~\cite{Go-Review, Orbital-Hall-Phonon, Orbital-Rashba}.

Lately, the OHE in two-dimensional (2D) materials has received a great deal of attention \cite{OHE_VHE_PRB_Imaging, Us4, Us1, Us2, Us3, Cysne-Bhowal-Vignale-Rappoport, OpticallyControlOrbitronic, OHE_Bhowal_1, OHE_Bhowal-Vignale, OrbitalPhotoCurrent, OrbitalPhotoCurrent-2}. Theoretical calculations predict the existence of orbital-textures in some 2D materials, which can give rise to the OHE \cite{BoropheneTexture-DFT,Us1, Luis-PRL,Us2}. They have been observed in insulating TMDs \cite{Orbital-Texture-Exp-TMD-1, Orbital-Texture-Exp-TMD-2}, where OHE plateaus are predicted \cite{Us1, Us2, Us3, Cysne-Bhowal-Vignale-Rappoport}. Furthermore, it is possible to attribute an orbital Chern number to this insulating phase \cite{Us3, Cysne-Bhowal-Vignale-Rappoport}, indicating a connection with nontrivial topology.

 \begin{figure*}
 \centering
    \includegraphics[width=0.31\linewidth,clip]{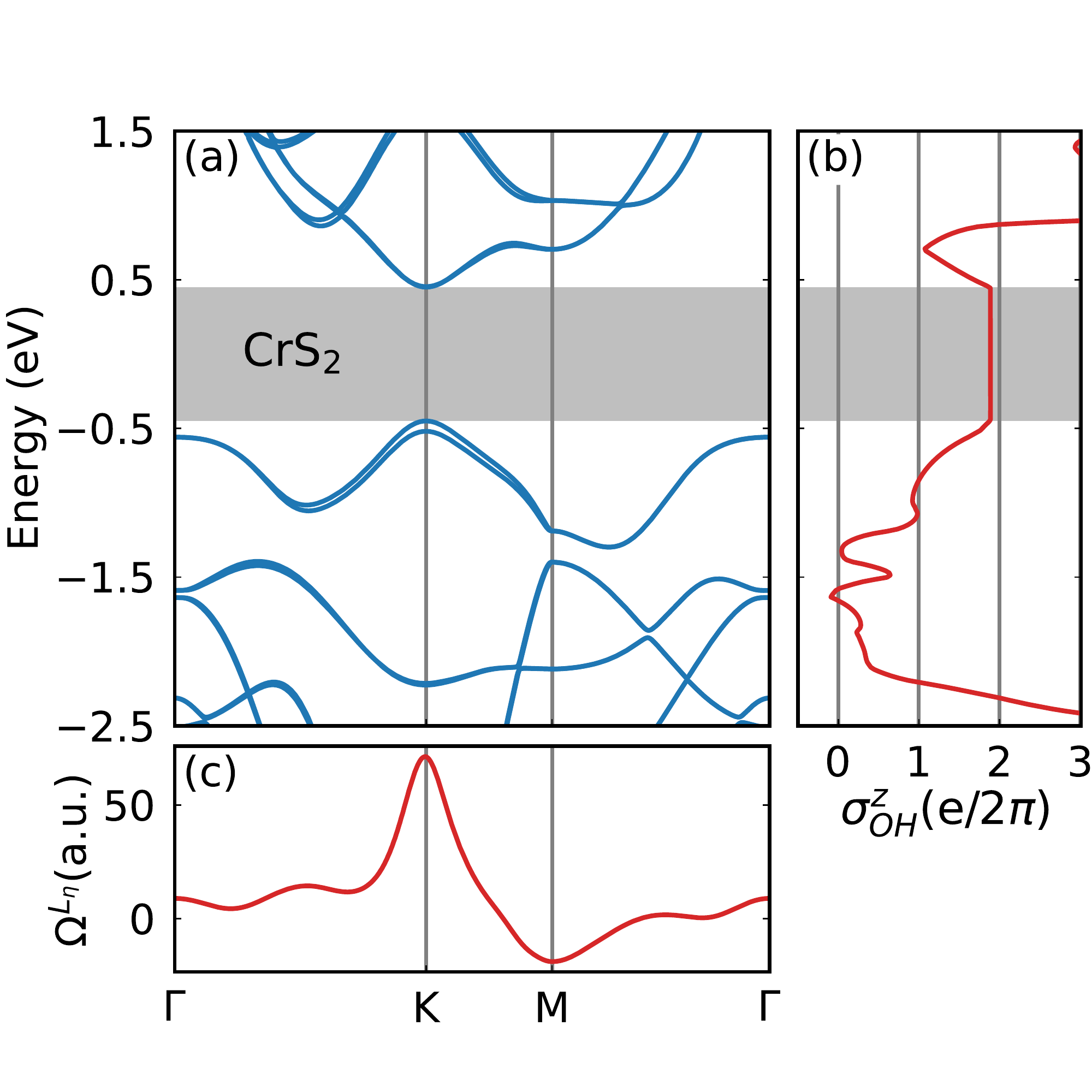}
    \hspace{1em}
    \includegraphics[width=0.31\linewidth,clip]{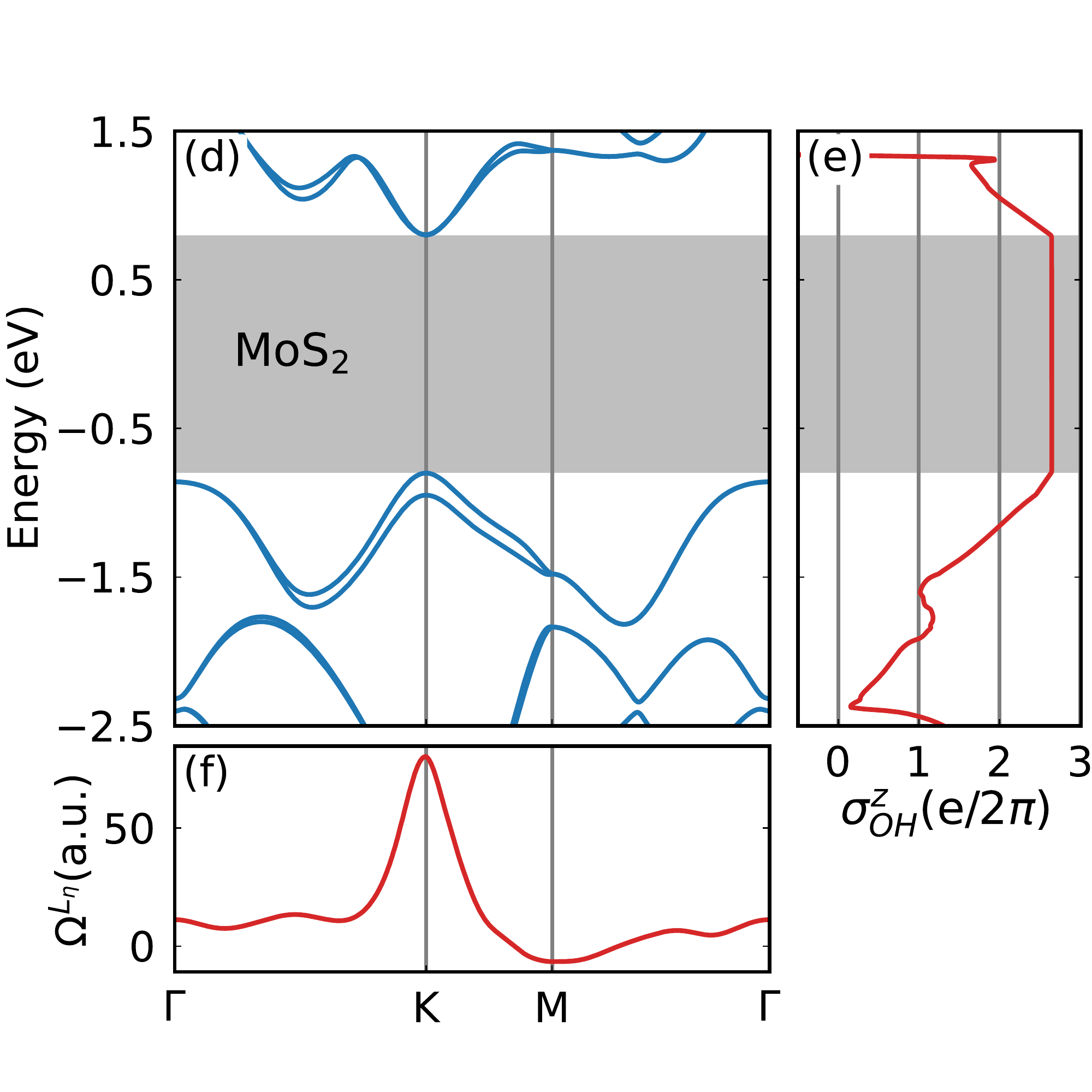}
    \hspace{1em}
    \includegraphics[width=0.31\linewidth,clip]{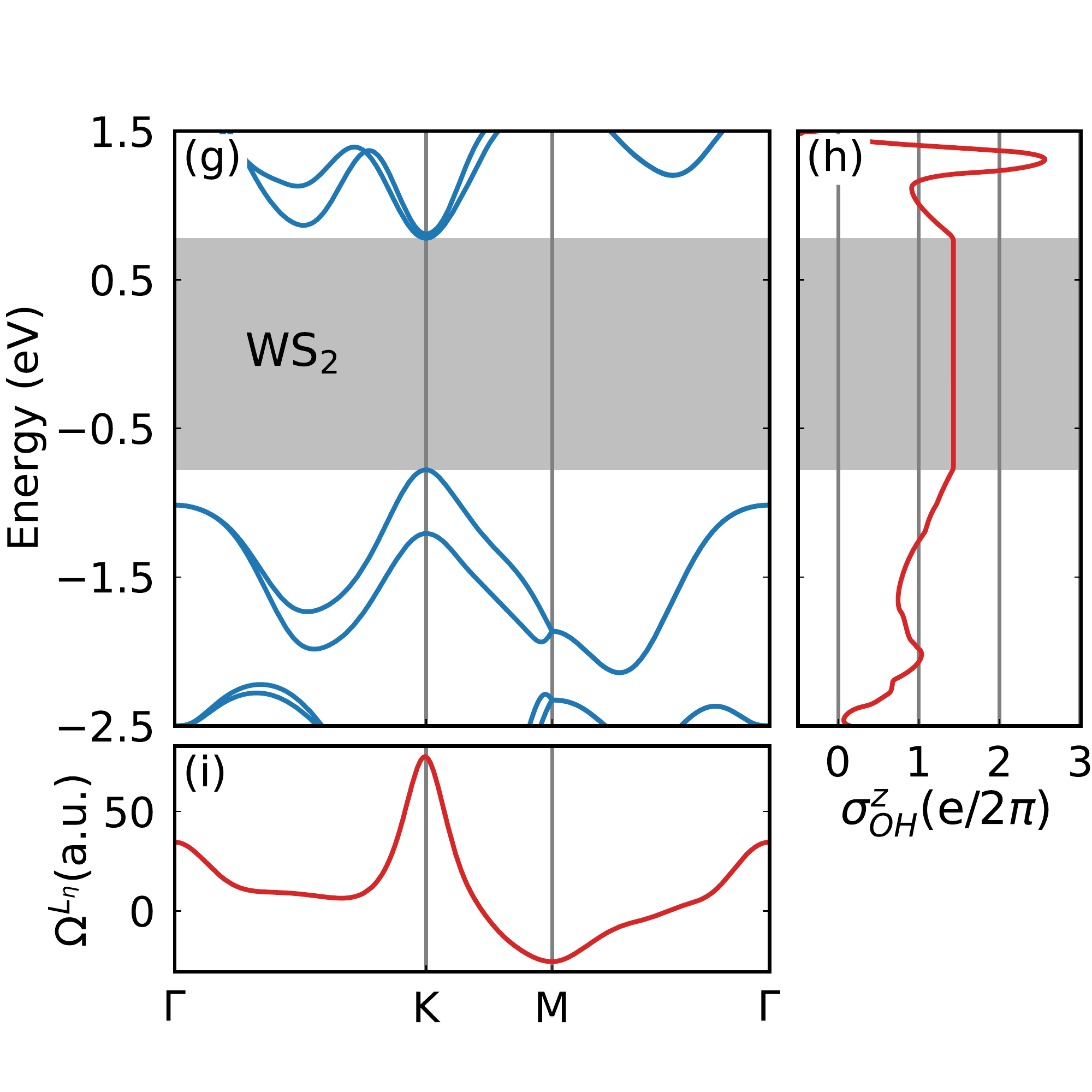}

\caption
{2H-TMD monolayers of CrS$_{2}$, MoS$_{2}$ and WS$_{2}$ fully relativistic band structures [(a), (d), and (g)] and orbital Berry curvatures [(c), (f), and (i)], calculated along high-symmetry directions of the 2D Brillouin zone. Panels (b), (e), and (h) display the corresponding OH conductivities calculated as function of energy.}

\label{fig1}
\end{figure*}

The topological nature of layered TMDs is mostly focused on the distorted structural phase 1T' that hosts topological insulators \cite{Tprime_Topology-1, Tprime_Topology-3,Giustino2020}. Less is known about the topology of octahedral (1T) and trigonal prismatic (2H) structures, where most monolayers exhibit an insulating character \cite{Yazyev2017}. MoS$_2$ is an archetype of a 2H-TMD that exhibits a large energy band gap. Their zigzag nanoribbons display metallic edge-states that cross the bulk gap \cite{Edgestates-TMD-1, Edgestates-TMD-2}, suggesting an underlying topology, despite being trivial with respect to the $\mathbb{Z}_2$ index \cite{Fabian-PseudohelicalEdgeStates}. 

Refs. \cite{Monalyer-TMD-corner-states, Bilayer-TMD-corner-states} have revealed that some 2H-TMDs are 2D higher-order topological insulators (HOTIs), not previously identified by the SHE signature \cite{hoti}. Triangular nanoflakes with armchair edges present in-gap corner states with fractional charge, protected by $C_3$ symmetry. Besides being one of the striking features of a 2D-HOTI \cite{Slager2015,HOTI-Bernevig, HOTI-Bernevig-2, HOTI-Rotation, HOTI-Rotation-2,hoti}, they explain the presence of metallic edge states in the zigzag edges and connect them to the topology of the 2D material. Their topological bulk polarization is 
perpendicular to the zigzag edges, leading to charge accumulation and metallic edge states.

Here, we use DFT and linear-response calculations to study the interplay between the OHE and HOTI phases in monolayers of TMDs respecting the rotation symmetry $C_3$.  We uncover that centrosymmetric 1T-TMDs can also be HOTIs. We then correlate the appearance of a HOTI phase in the two different structures with an orbital Hall insulating phase. We discuss the existence of a pseudo-time reversal symmetry, which originates from the crystalline symmetries of the lattice. In analogy with the photonic quantum spin Hall effect, eigenstates of the orbital angular momentum play the role of pseudo-spins. They are also eigenstates of the $C_3$ rotation operator that protects the HOTI. Because of the orbital nature of the pseudo-spins, the higher-order topological phase can be witnessed by the OHE. We use a well-known low energy model for HOTIs \cite{Schindler2018} and the three $d$-orbitals model for 2H-TMDs \cite{ThreeBandTMD} to explicitly show this connection \cite{SuplementaryMaterial}. 

{\it OHE calculations for TMDs in the 2H and 1T structural phases: \ }  
We begin by presenting the band structure and the OHE for the two families. For that purpose, we performed DFT calculations for 1T and 2H TMD monolayers (MX$_2$, where M is a transition metal and X is a chalcogen). The TMD structures were obtained from the C2DB database \cite{C2DB}. We also adopted their criteria for dynamic (phonons) and thermodynamic stability. We fully optimize the structural parameters to obtain the wavefunctions. Then, we construct a PAO Hamiltonian using the pseudo-atomic-orbital (PAO) projection method \cite{PAO1,PAO2, PAO3, PAO4} for each compound. This method is implemented in the \textsc{paoflow} code \cite{PAO5,PAO6}; for technical details see supplementary material (SM).

Once the PAO Hamiltonian $\mathcal{H}_{\text{PAO}} ({\bf k})$ is built, we calculate the spin Hall (SH) and orbital Hall (OH) conductivities to linear order on the external electric field~\cite{Us1, Us2, Us3} 
\begin{eqnarray}
\sigma^{\eta}_{OH(SH)}=\frac{e}{(2\pi)^2}\sum_{n} \int_{BZ} d^2k f_{n {\bf k}}~\Omega_{n, {\bf k}}^{X_{\eta}},
\label{conductivity}
\end{eqnarray}
where $\sigma^{\eta}_{OH(SH)}$ is the OH (SH) DC conductivity with polarization along the $\eta$-direction, $f_{n {\bf k}}$ is the Fermi-Dirac distribution and $\Omega_{n, {\bf k}}^{X_{\eta}}$  is the angular momentum projected Berry curvature in the intra-atomic approximation \cite{Us1,Us2,Us3} (see SM \cite{SuplementaryMaterial}). 

\begin{figure*}
 \centering
    \includegraphics[width=0.31\linewidth,clip]{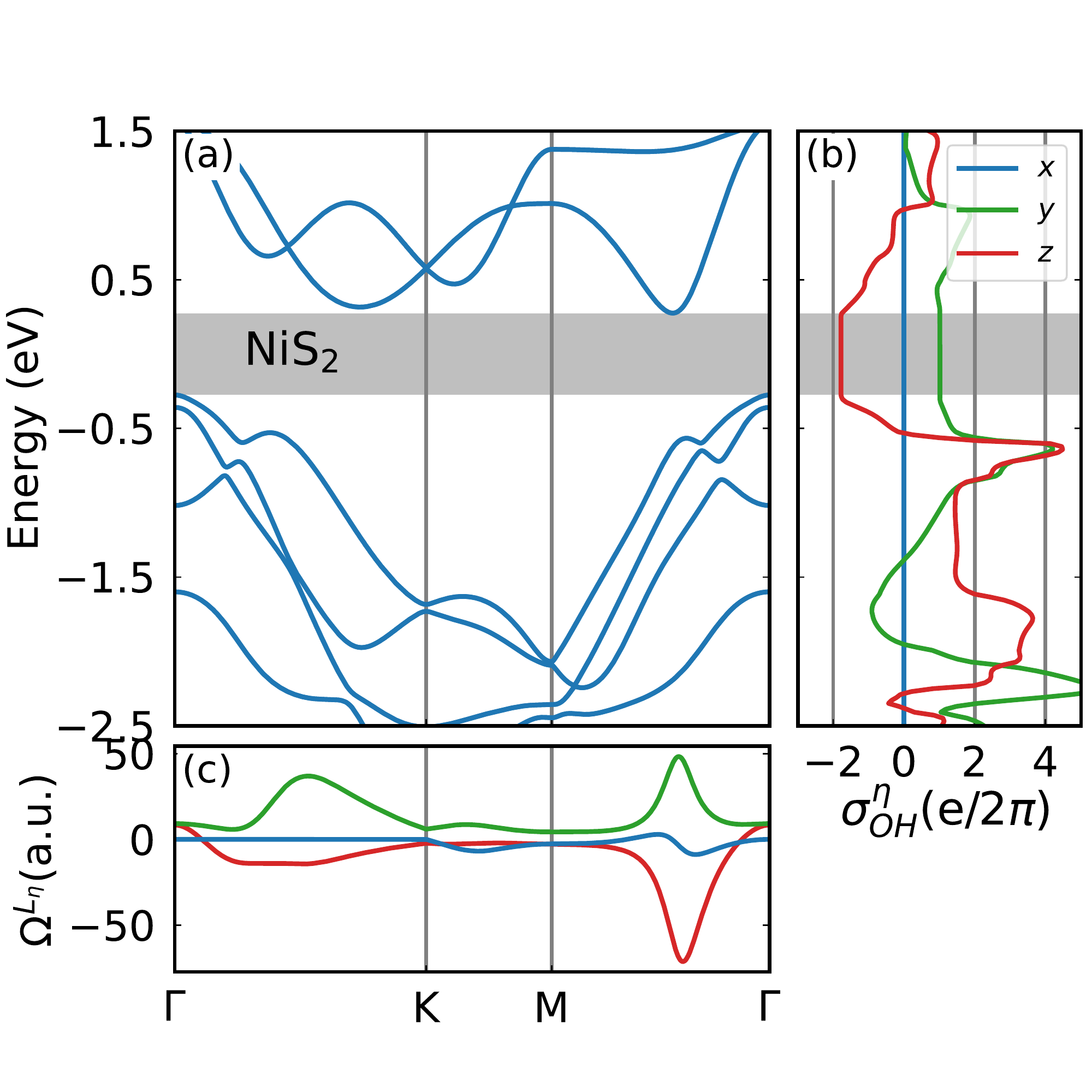}
    \hspace{1em}
    \includegraphics[width=0.31\linewidth,clip]{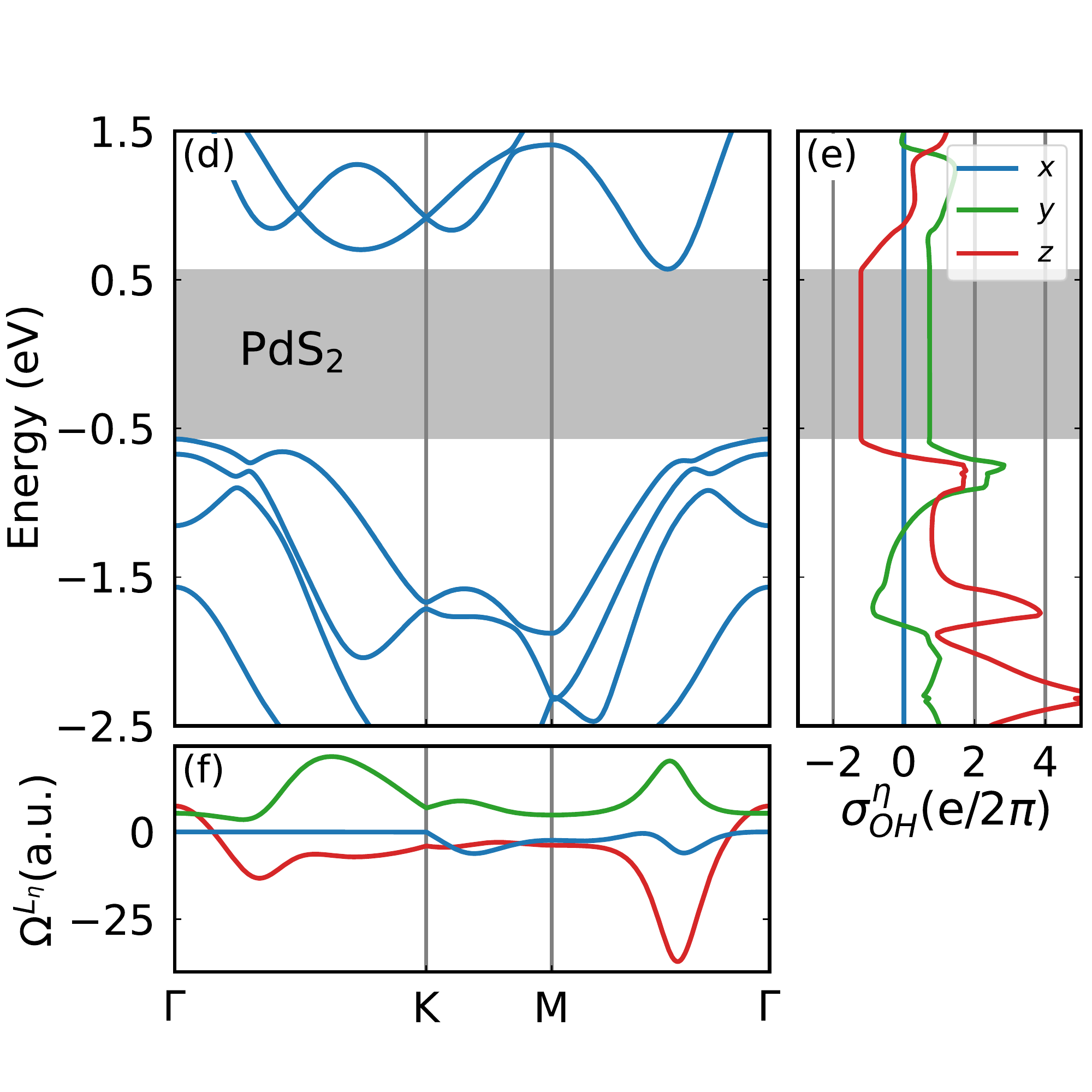}
    \hspace{1em}
    \includegraphics[width=0.31\linewidth,clip]{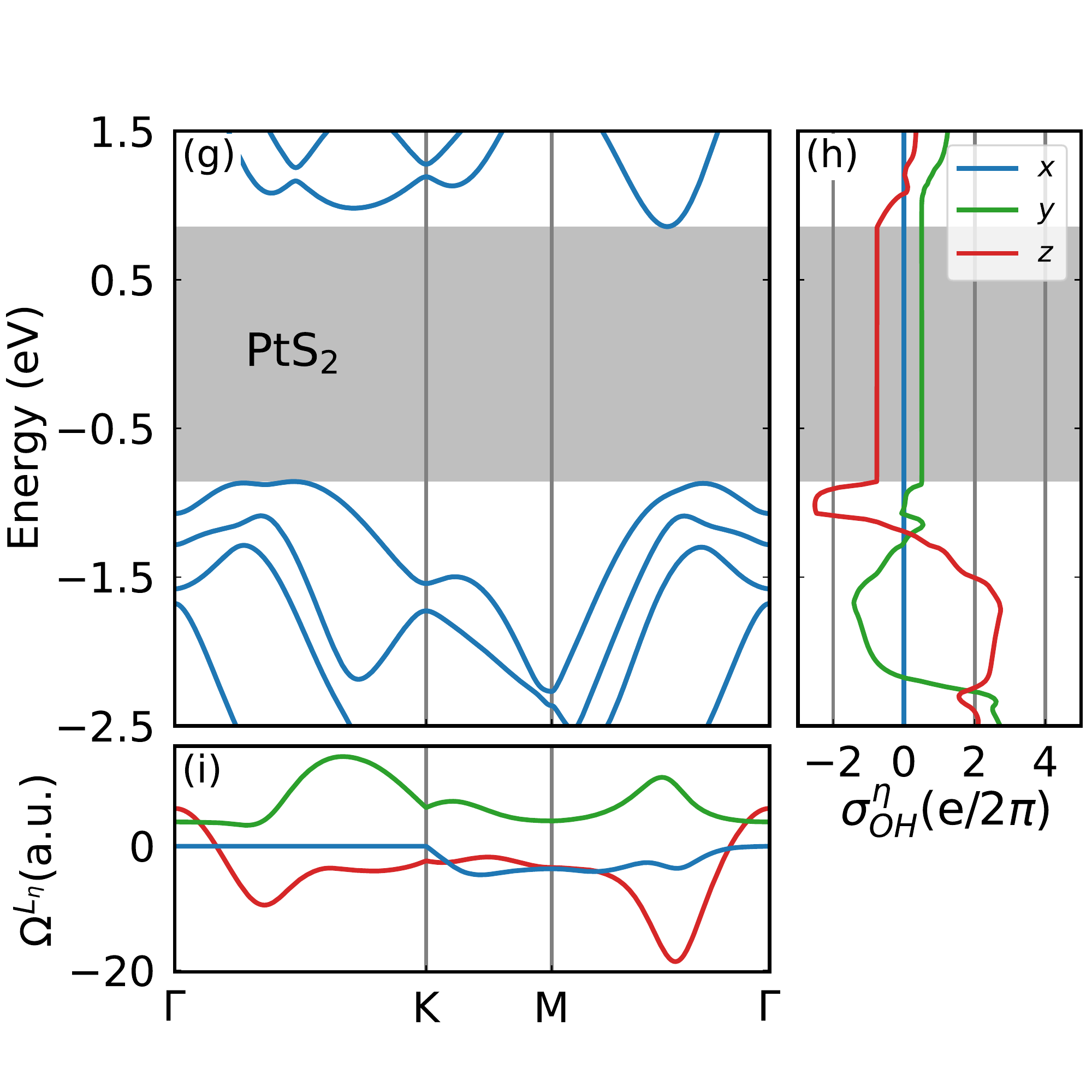}

\caption
{1T-TMD monolayers of NiS$_{2}$, PdS$_{2}$ and PtS$_{2}$ fully relativistic band structures [(a),(d) and (g)] and orbital Berry curvatures [(c), (f) and (i)], calculated along high-symmetry directions of the 2D Brillouin zone. Panels (b), (e) and (h) display the corresponding OH conductivities calculated as function of energy.}

\label{fig2}
\end{figure*}

Figures \ref{fig1} and \ref{fig2} show the fully relativistic band structures, OH conductivities and the orbital Berry curvatures calculated for different TMD monolayers. In Fig. \ref{fig1} we depict results for 2H-TMDs  CrS$_2$, MoS$_2$ and WS$_2$. We note that within their insulating gaps these systems exhibit OH conductivity (OHC) plateaus with values $\sigma^{z}_{OH}=$ 1.89, 2.65, and 1.43 in units of $(\frac{e}{2\pi})$, respectively. Conversely,  $\sigma^{x}_{OH}$ and $\sigma^{y}_{OH}$ vanish for all three systems. Their corresponding orbital Berry curvatures are  similar. They are peaked around the K point with similar maxima for the three systems, giving a large positive contribution to the OHC. Around M, the Berry curvature becomes negative and have different values for the three compounds, resulting in the plateau variations obtained for the OHCs when we go from Cr to W. Since the top valence bands of the TMDs have predominantly $d$-character, the chalcogen element does not significantly affect the OHC. Table 1 of the SM \cite{SuplementaryMaterial} shows that the in-gap OHC does not change much when S is replaced by Se or Te.

Figure \ref{fig2} presents results for the 1T-TMDs monolayers NiS$_2$, PdS$_2$ and PtS$_2$. Differently from the 2H-TMDs, the OHCs for the 1T-TMDs with  $y$ polarization are not zero, and thus contribute to the entire OHC. We follow ref. \cite{Jose_PRL_modSigma} and define the absolute value of the OHC as $|\sigma_{OH}| = \sqrt{(\sigma^x_{OH})^2 + (\sigma^y_{OH})^2 + (\sigma^z_{OH})^2}$.  Within each energy band-gap, we clearly see that the dominant contribution to $|\sigma_{OH}|$ is $\sigma^z_{OH}$ for all three systems, and the OHC plateau reduces as we move from Ni to Pt.  

Table \ref{table_TTMD} summarizes the main findings depicted in Fig. \ref{fig2}. It is noteworthy that both 2H and 1T TMD monolayers exhibit finite OH conductivity plateaus within their insulating energy gaps. This shows that the OHE in 2D materials is not constrained by spatial inversion symmetry and can also appear in centrosymmetric monolayers. Our results are inline with recent predictions showing that centrosymmetric 2H TMD bilayers  also display OHC plateaus \cite{Us2, Us3, Cysne-Bhowal-Vignale-Rappoport}.  We note that $|\sigma_{SH}|=0$ inside the band gap for both sets, which is consistent with the fact that TMDs do not exhibit a QSHE in the structural phases studied here. 
\begin{table}
	\centering
		\centering
	\begin{tabular}{|c c | c c | c c | c c | c c | c c | c |} 
		\hline
		TMD (Z$_\text{M}$)& & $\sigma^{x}_{OH}$ & &  $\sigma^{y}_{OH}$ & &  $\sigma^{z}_{OH}$ & & $|\sigma_{OH}|$ & &  $E_g (\text{eV})$ & & $\mathbb{Z}_4$    \\ [0.5ex] 
		\hline
		\hline
		NiS$_2$ (28) & & $0$ & & $1.02$ & & $-1.78$ & & $2.05$ & & $0.54$ & & 2   \\
		\hline
	        PdS$_2$ (46)  & & $0$ & & $0.73$ & & $-1.22$ & & $1.85$ & & $1.14$ & & 2  \\
		\hline
	        PtS$_2$ (78) \cite{1T-PtS2-Syntesis} & & $0$ & & $0.51$ & & $-0.77$ & & $0.92$ & &  $1.72$ & & 2  \\
		\hline
	\end{tabular}
	\caption{
	Main characteristics of the 1T-TMD monolayers insulating phases of NiS$_2$, PdS$_2$, and PtS$_2$. $Z_{\text{M}}$ is the atomic number of the constituent transition metal atom. The columns $\sigma^{x}_{OH}$, $\sigma^{y}_{OH}$, $\sigma^{z}_{OH}$ and $|\sigma_{OH}|$  show their OHE in-gap values in units of $e/(2\pi)$. $E_g$ is the energy band gap and the last column shows the values of the topological invariant $\mathbb{Z}_4$.}
	\label{table_TTMD}
	
\end{table}
In the SM, we include tables containing several 2H and 1T semiconducting TMD monolayers \cite{SuplementaryMaterial}. We present their electronic band structures and orbital-weighted Berry curvatures for the valence bands and discuss some of their features.

{\it HOTI phase:}
We now proceed to the characterization of the HOTI phases. Recent works showed that triangular nanoflakes of 2H$-$TMD monolayers with armchair edges present in-gap corner states with fractional charge $(-\frac{1}{3}|e|)$, protected by $C_3$ symmetry~\cite{Monalyer-TMD-corner-states, Bilayer-TMD-corner-states}. We begin our analysis by calculating the topological indicators for this TMD family. 

For non-centrosymmetric materials, the HOTI phase is protected by a $C_n$ rotation symmetry. It can be identified by the symmetry representations of the occupied energy bands at special high-symmetry points (HSP) of the first Brillouin zone (BZ) \cite{HOTI-Rotation-2}. For $C_3$ rotation symmetry, we take $\left[K_p^{(3)}\right] = \#K_{p}^{(3)} - \#\Gamma_{p}^{(3)}$, where $\#K_{p}^{(3)}$ and $\#\Gamma_{p}^{(3)}$ represent the number of occupied bands with symmetry eigenvalue $e^{2\pi i(p-1)/3}$ (for $p=1, 2, 3$) at the K$-$ and $\Gamma-$ high symmetry points, respectively. The final topological indicator $\chi^{(3)}$ and  corner charge $Q_{\rm c}^{(3)}$ are given by
\begin{equation}
    \chi^{(3)} = \left(\left[K_1^{(3)}\right],\left[K_2^{(3)}\right]\right), ~~~
    Q_{\rm c}^{(3)} = \frac{e}{3} \left[K_2^{(3)}\right] {\rm mod}\;e, 
\end{equation}
where $e$ is the elemental charge. We use the software IrRep~\cite{IRAOLA2022108226} to calculate the symmetry eigenvalues of the occupied DFT energy bands. With them, we calculate the topological indicator and the corner charge with the expressions above.  The three 2H-TMD monolayers presented here have the same topological indicator $\chi^{(3)}=[-1,2]$  and corner charge $Q_{\rm c}^{(3)}=2e/3$. A table for several 2H-TMDs is included in the SM\cite{SuplementaryMaterial}. 

\begin{figure}
 \centering
    \includegraphics[width=0.97\columnwidth]{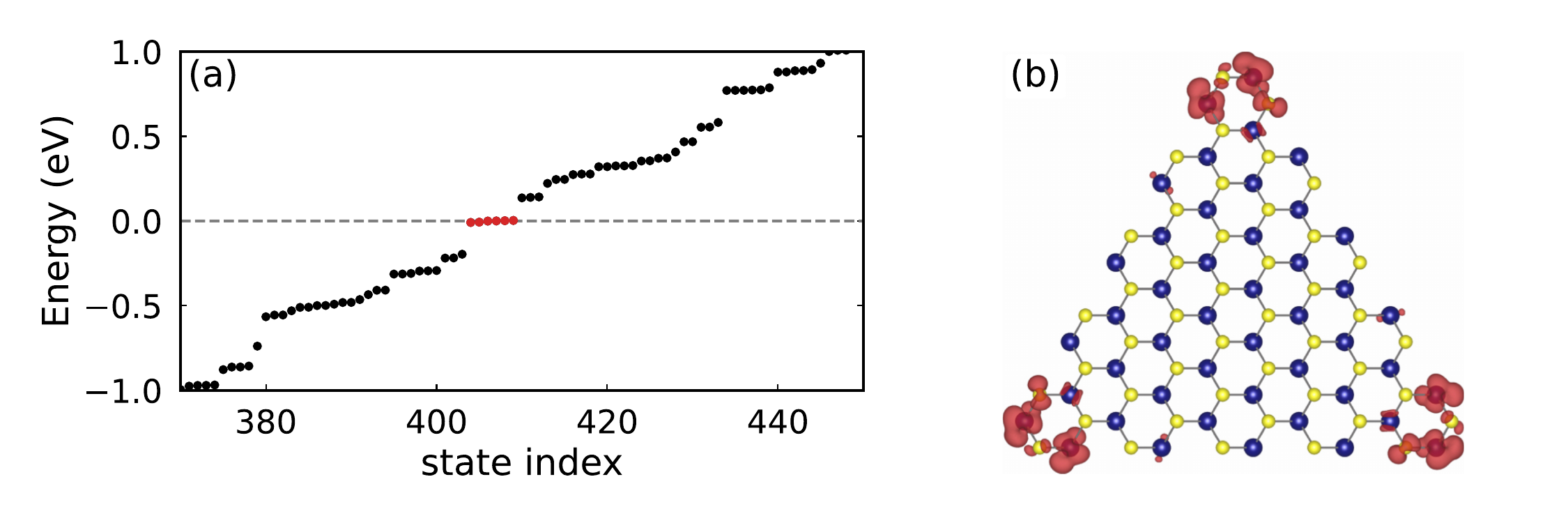}
\caption
{Monolayer MoS$_2$ nano-flake (0D) geometry. (a) Fully relativistic calculation eigenvalues. The corner states are highlighted in red. (b)  Real space projection of the eigenfunctions for the corner states highlighted in panel (a). Isosurface value of 0.003 $\mathrm{e}$\AA{}$^{-3}$.}
\label{fig3}
\end{figure}
To complement the analysis based on the eigenstates of the rotation operator, we use DFT to examine 2H-TMD triangular flakes  with armchair edges and confirm the presence of in-gap corner states, as shown in figure \ref{fig3}. 
2H-TMDs also display an electronic dipole $\mathbf{P}=(\frac{1}{3},\frac{2}{3}$), which is perpendicular to the zigzag direction \cite{Monalyer-TMD-corner-states}.  As a result, if the system is cut in the zigzag direction, there is charge accumulation at the edges, leading to metallic edge states. 

Differently from the 2H-TMDs, 1T-TMDs are HOTIs protected by inversion symmetry. Hence, they are characterized by the $\mathbb{Z}_4$ indicator, which can be calculated from the inversion parities of occupied bands \cite{Kruthoff2017,Z4}: $    \mathbb{Z}_4 = \sum_{k_i \in {\rm TRIMs}} n_{-}(k_i) \;\;{\rm mod}\;4, $
 where $n_{-}(k_i)$ is the number of odd parity occupied Kramer pairs at the time reversal invariant momenta (TRIM) points $k_i$ in the BZ.

\begin{figure}
 \centering
    \includegraphics[width=0.97\columnwidth]{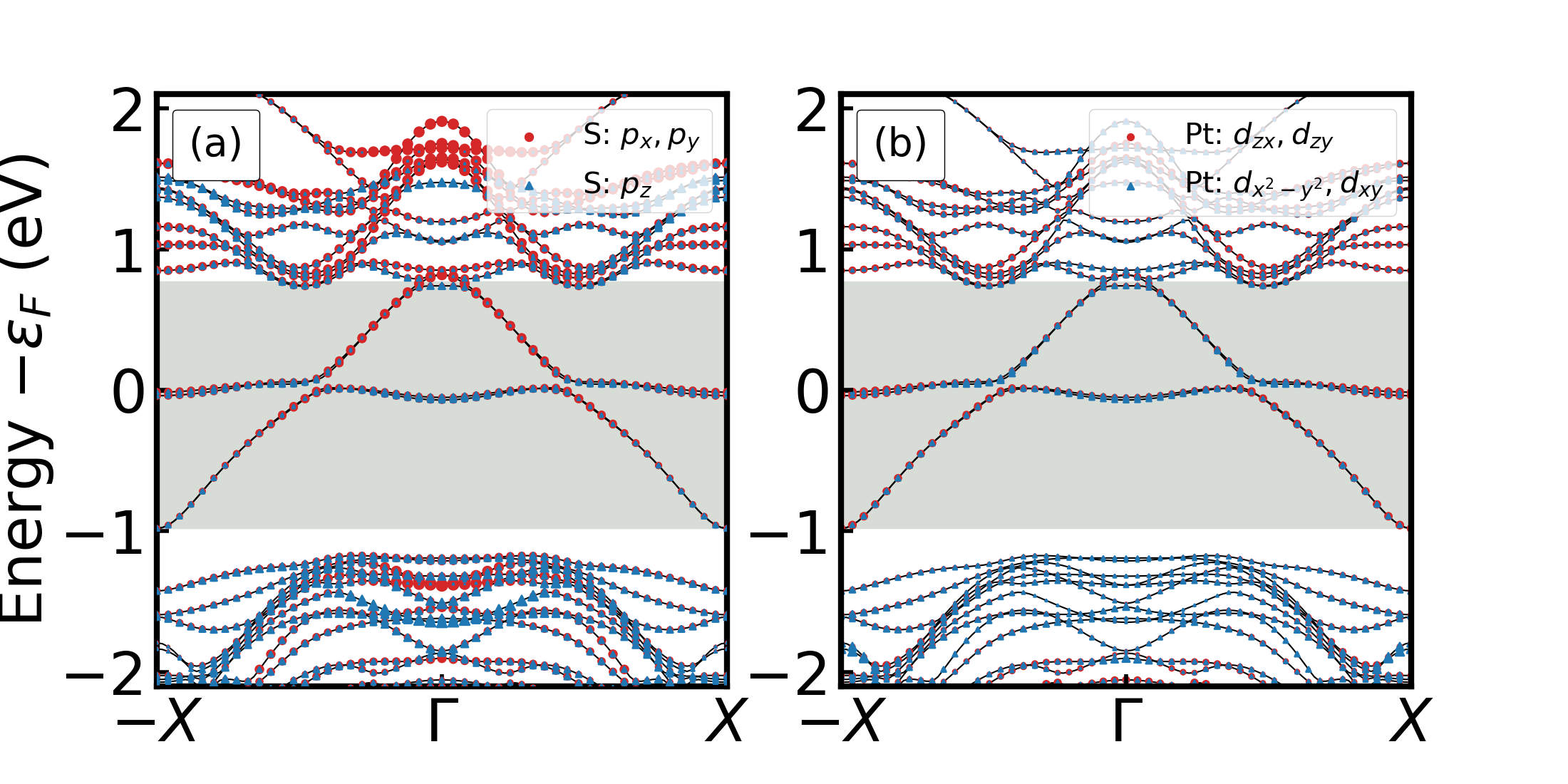}
\caption
{Orbital projected band structures of 18.6 \AA-wide zigzag nanoribbon of a PtS$_2$ monolayer calculated with the PAOFLOW Hamiltonian. The shaded area and colored markers represent the bulk bandgap and  orbital projections, respectively. (a) Contributions from the $p_x,p_y$ (red circles) and $p_z$ (blue triangles) orbitals from S atoms. (b) Contributions from the $d_{xz},d_{yz}$ (red circles) and $d_{x^2-y^2},d_{xy}$ (blue triangles) orbitals of Pt atoms.
}
\label{fig4}
\end{figure}

The index $\mathbb{Z}_4=2$ warrants that 1T-TMDs also present conducting edge-states capable of carrying OAM currents. This can be confirmed from the energy bands of a PtS$_2$ nanoribbon with zigzag edges portrayed in figure \ref{fig4}. This figure also highlights  the orbital projection of the edge states. Because of the inversion symmetry, Bloch states of 1T-TMDs do not exhibit net OAM but can still display OHE \cite{OrbitalTexture}. As a result, the nanoribbon bands do not have any OAM polarization. This contrasts with the 2H-TMD nanoribbons that have well-defined orbital-polarized edge states. 

{\it Discussion:} We have shown that a large set of insulating 2D materials exhibits sizeable OHE coexisting with a HOTI phase. This shows that higher-order topology may allow in-gap conducting edge states that can transport orbital angular momentum in the orbital Hall insulating phase. The bulk polarization perpendicular to the edges of a 2D HOTI yields to charge accumulation at the edges. As a result, these systems can have conductive edge states within their bulk energy band gaps when cut along certain directions.

To connect the OHE with the HOTI phase, it is worth recalling how the quantum spin Hall effect (QSHE) is emulated in photonic crystals. For the appearance of a $\mathbb{Z}_2$ topological phase, one needs Kramers degenerate pairs. However, bosonic systems do not possess half-integer spins. Thus, under the action of time-reversal symmetry operation (${\mathcal {T}}$), they transform as ${\mathcal {T}}^2=1$, whereas for fermionic systems,  ${\mathcal {T}}^2=-1$ \cite{RMOPTopologicalPhotonics}. To engineer the photonic QSHE, one may construct a Hamiltonian that is invariant under inversion and a pseudo-time-reversal symmetry ($\mathcal{T}_p$) so that ${\mathcal{T}_p}^2=-1$. This is achieved with a lattice of dielectric cylinders, which work as artificial atoms exhibiting electronic orbital-like shapes that produce the photonic bands \cite{SchemetopologiacalPhotonic}. The pseudo-time-reversal operator originates from a combination of lattice symmetry operations in such a way that the pseudo-spins are eigenstates of $L_z$ such as $p_{\pm}=p_x\pm i p_y$  and $d_{\pm}= d_{xy}\pm i d_{x^2-y^2}$ \cite{PseudoTRAcoustic,SchemetopologiacalPhotonic}. 

To identify these pseudo-spins in the TMDs, we can use general symmetry considerations regarding 2D materials with $3$-fold rotational symmetry. The crystal field in low-dimensional systems leads to large splittings between orbitals, inducing the formation of energy gaps. However, the rotational symmetry also imposes constraints on the energy states. These have to be also eigenstates of the rotation operator at the high symmetry points.

For 2H-TMDs, the conduction and valence bands at $K $ and $K'$ are mainly composed of $d_{z^2}$ and $d_{\pm}$ orbitals, which are eigenstates of $L_z$. This is a consequence of the $D_{3h}$ point group symmetry of the crystal: the $d_{z^2}$ orbitals belong to the unidimensional irreducible representation $A_1$, while the $d_{xy}$ and $d_{x^2-y^2}$ belong to $E^\prime$. Since the representation $E^\prime$ is 2-dimensional, the linear combinations $d_{\pm}$ can be treated as pseudo-spins that transform under a pseudo-time-reversal symmetry operator related to the rotation operators.  In the SM \cite{SuplementaryMaterial}, we show that the Chern number associated with $\mathcal{T}_p$ symmetry leads to the same result previously obtained \cite{Us3}.

In the case of the 1T TMDs, the identification of the pseudo-spins is more subtle. The $D_{3d}$ symmetry imposes that the $p$ orbitals of the X atoms and the $d$ orbitals from the M atom should be dominant for the bulk energy states near $\Gamma$. Our first-principles calculations \cite{SuplementaryMaterial} evince a strong energy splitting in $\Gamma$ between the $p_{x},p_{y}$ and the $p_z$ orbitals, as previously reported by Yao \textit{et al.} for PtSe${}_2$ \cite{PtSe2Spin-Layer}. The valence and conduction bands near the gap have a strong contribution of linear combinations of $p_x$ and $p_y$ orbitals of the X atom that form $p_{\pm}$, which transform as ${\mathcal{T}_p}^2 = -1$. They are also composed of $d_{\pm}$. The sizable contribution of $d_{xz}$ and $d_{yz}$ orbitals explain $L_x$ and $L_y$ components of the OHE. When the system is cut into a ribbon, these orbitals participate in the formation of edge states, as shown in Fig. \ref{fig4}(b). Therefore, in contrast with 2H-TMDs, these edge states will be mainly composed of combinations of three pseudo-spins, one formed by the $p_x,p_y$ orbitals and two composed of the $d_{xz},d_{yz},d_{x^2 - y^2}$, and $d_{xy}$ orbitals.

In principle, similar to photonics systems, one could use these eigenstates of $L_z$ to emulate topological phases in fermionic materials: a system which is invariant under inversion and  ${\mathcal {T}}_p$, must have pseudo-spins forming Kramers' pairs. However, differently from bosons, fermions have half-integer spins, and two spin-degenerate states for each pseudo-spin. Therefore, if one tries to construct a fermionic system without spin-orbit coupling where the pseudo-spins emulate a quantum spin Hall insulator, the system has spin-degenerate pseudo-spins Kramers' pairs. This results in an even number of Kramers' pairs and the system cannot be indexed by a $\mathbb{Z}_2=1$, although it can be a HOTI.

To illustrate these ideas, we use the low energy Hamiltonian presented in Ref. \onlinecite{Schindler2018} to model HOTIs protected by $C_3$ rotation and inversion symmetries (see SM \cite{SuplementaryMaterial}).  This model consists of a block diagonal Hamiltonian containing basically the superposition of two copies of the Bernevig-Hughes-Zhang (BHZ) model. It is well known that each BHZ Hamiltonian presents a $\mathbb{Z}_2$ topological phase. Its eigenstates can also be written in terms of pseudo-spins that are eigenstates of $L_z$. Surprisingly, we show they present an orbital Hall plateau in their topological gap. When the two BHZ copies are taken into account, the system is not a topological insulator but, as shown in Ref. \onlinecite{Schindler2018}, it is a HOTI that has twice the number of edge states of the BHZ model. As expected, the HOTI still presents an orbital Hall plateau and the orbital current can be carried by the in-gap edge states.

To strengthen this link, we used another model to show the onset of a HOTI phase in systems without inversion symmetry, following the ideas presented in Ref. \onlinecite{Eck2022}.  We considered the simplified three orbital tight-binding Hamiltonian in a triangular lattice that describes the low-energy properties of 2H-TMDs. We begin with a case with inversion symmetry and orthogonality between the orbitals in different representations. Under this condition, a strong spin-orbit coupling opens a gap in the system, leading to a trivial insulator phase with a vanishing OHE and the absence of in-gap edge states. From this, we identify that if inversion symmetry is broken and hopping between orthogonal orbitals is allowed, as in the case of 2H-TMD, there is a topological transition to a HOTI that presents zigzag metallic edge states and a large OH plateau, which is independent of the SOC \cite{SuplementaryMaterial}.

{\it Conclusions:} We employed  DFT and linear response transport calculations to study the interplay between the orbital Hall effect and higher-order topological phases. We analyzed all stable 2H or 1T monolayer TMDs and found that they are HOTIs, protected by either $C_3$ rotation symmetry (2H) or inversion symmetry (1T). Simultaneously, they all display a plateau in the orbital Hall conductivity inside the band gap. 

Recent works start to uncover the role of orbital hybridization in HOTIs and the advent of orbital effects \cite{Mazanov2022,Eck2022,Gliozzi2022}. Here, we connect the HOTI phase to the existence of pseudo-time reversal operators and associated pseudo-spinors. As these pseudo-spinors are eigenstates of the orbital angular momentum, HOTI phases can generate OHE. More importantly, HOTIs with edges that are perpendicular to their bulk polarization present in-gap metallic edge states that can carry the orbital angular momentum in the orbital Hall insulating phase. This can be employed for efficient orbital current injection in novel  spin-orbitronics devices.  Furthermore, the OHE in 2D HOTIs may be used in machine learning strategies for spotting potentially useful materials for orbitronic applications \cite{ML_TIs,ML_2D,Mat_bigdata}.

\begin{acknowledgments}
	We acknowledge CNPq/Brazil, CAPES/Brazil, FAPERJ/Brazil, INCT Nanocarbono and INCT Materials Informatics for financial support. TGR acknowledges funding from Fundação para a Ciência e a Tecnologia and Instituto de Telecomunicações - grant number UID/50008/2020 in the framework of the project Sym-Break. M. C. acknowledges CNPq (Grant No. 317320/2021-1) FAPERJ/Brazil (Grant No. E26/200.240/2023) and the National Laboratory for Scientific Computing (LNCC/MCTI, Brazil) for providing HPC resources. B.F. acknowledges funding from FAPESP/Brazil under grant no. 2019/04527-0 and the Brazilian Nanotechnology National Laboratory (LNNano/CNPEM, Brazil) for computational resources. A.F. acknowledges funding from FAPESP/Brazil under grant no. 2017/02317-2. L.M.C acknowledges the funding from the ECONWHET project, reference PID2019-106684GB-I00, funded by MCIN/ AEI /10.13039/501100011033/ and by "ERDF A way of making Europe". ICN2 is funded by the CERCA Programme/Generalitat de Catalunya and supported by the Severo Ochoa Centres of Excellence program, funded by the Spanish Research Agency (Grant No. SEV-2017-0706).
\end{acknowledgments}

\bibliographystyle{apsrev}

\onecolumngrid
\
\renewcommand{\theequation}{S\arabic{equation}}
\renewcommand{\thefigure}{S\arabic{figure}}

\newpage
{\center \large \bf Supplementary material for ``Connecting Higher-Order Topology with the Orbital Hall Effect in Monolayers of Transition Metal Dichalcogenides''}

\section{Calculation of the orbital Hall conductivity: computational details}

The density functional theory (DFT) calculations~\cite{DFT1,DFT2} were performed with the plane-wave-based code \textsc{Quantum Espresso}~\cite{QE-2017}. The exchange and correlation potential were treated within the generalized gradient approximation (GGA)~\cite{PBE}. The ionic cores were described with fully relativistic projected augmented wave (PAW) potentials~\cite{PAW}. The wavefunctions and charge density energy were 40\% larger than the pslibrary recommended \cite{pslibrary} value. Our self-consistent calculations (SCF) were performed with a linear density of $\bf k$-points of 12.0/\AA$^{-1}$, and to avoid spurious interactions a minimum of 15 \AA{} of vacuum is used.

We constructed an effective tight-binding Hamiltonian from our DFT calculations using the pseudo atomic orbital projection (PAO) method~\cite{PAO1,PAO3} as implemented in the \textsc{paoflow} code\cite{PAO5,PAO6}. The PAO method consists of projecting the several thousand plane-waves DFT Kohn-Sham orbitals onto the compact subspace spanned by the pseudo atomic orbitals, which are naturally built-in into the PAW potentials. This procedure significantly reduces the computational cost of performing large integration's. We have used this method to investigate topological properties \cite{Costa2019,Costa2018}, spin dynamics \cite{adatoms,fegete}, transport properties \cite{hoti,cri3-graphene} and others. The orbital Hall conductivity calculations were performed with a reciprocal space sampling 10 times larger than the DFT-SCF calculations.

\section{Orbital Weighted Berry Curvature and Electronic Spectra} 

The orbital-weighted Berry curvature $\Omega_{n, {\bf k}}^{X_{\eta}}$ is given  by
\begin{eqnarray}
\Omega_{n, {\bf k}}^{X_{\eta}}= 2\hbar\sum_{m\neq n}\text{Im} \Bigg[ \frac{\langle u_{n,{\bf k}}\big|j_{y,{\bf k}}^{X_{\eta}}\big|u_{m,{\bf k}}\rangle \langle u_{m,{\bf k}}\big|v_x({\bf k})\big|u_{n,{\bf k}}\rangle}{(E_{n,{\bf k}}-E_{m,{\bf k}}+i0^+)^2}\Bigg].\label{Kubo2}
\end{eqnarray}
The velocity operators are given by $v_{x(y)}({\bf k})=\hbar^{-1}\partial \mathcal{H}_{\text{PAO}} ({\bf k})/\partial k_{x(y)}$, and  $\big|u_{n(m),{\bf k}}\rangle$ is the periodic part of Bloch wave function with energy $E_{n(m),{\bf k}}$. The orbital (spin) current polarized in $\eta$-direction ($\eta=x, y, z$) is defined by $j_{y,{\bf k}}^{X_{\eta}}=\left(X_{\eta}v_{y}({\bf k})+v_{y}({\bf k})X_{\eta} \right)/2$, where $X_{\eta}=\hat{\ell}_{\eta} (\hat{s}_{\eta})$ is the $\eta$-component of OAM (spin) operator. Here, we use the intra-atomic approximation for the OAM operator that gives a reliable description of the OHE for TMDs \cite{Us3, Cysne-Bhowal-Vignale-Rappoport}.

The orbital-weighted Berry curvature of the valence-band is defined as the sum over the occupied valence-band states of the individual orbital-weighted curvatures given by 

\begin{eqnarray}
\Omega^{L_\eta}({\bf k})=\sum_{n\in \text{val}}\Omega^{L_\eta}_{n,{\bf k}}.
\label{OWBerryValenceBand}
\end{eqnarray}
At zero temperature and for Fermi energies within the TMD band gap, we note that the orbital-Hall-conductivity plateau $\sigma^{\eta}_{OH}$, associated to orbital angular momentum component $L_{\eta}$ ($\eta=x,y,z$), is given by 
\begin{equation}
    \sigma^{\eta}_{OH}=\frac{e}{(2\pi)^2}\int_{BZ} d^2k~\Omega^{L_\eta}({\bf k}),
\label{conductivity}
\end{equation}
where the integral is over the Brillouin zone (BZ). Here, we follow Ref. \citenum{Jose_PRL_modSigma} and define $|\sigma_{OH}|=\sqrt{(\sigma^{x}_{OH})^2+(\sigma^{y}_{OH})^2+(\sigma^{z}_{OH})^2}$.

In the following sections we present our DFT-calculated results of $\sigma^{\eta}_{OH}$ and $|\sigma_{OH}|$ for several TMD monoloyers in the 2H and 1T structural phases, together with their corresponding electronic energy bands.  
 
\section{2H-TMDs}

Table \ref{table_2HTMD_Sup} extends the table I presented in the main text. It includes all insulating TMD monolayers that naturally stabilize in the 2H structural phase. They are listed here in decreasing order of their orbital Hall conductivity plateaus.
\begin{table*}[h!]

	\begin{tabular}{||c c || c c | c c | c c | c c | c c | c c | c c | c c | c ||} 
		\hline
		H-TMD (Z$_\text{M}$) & & $\sigma^{x}_{OH}$ & &  $\sigma^{y}_{OH}$ & &  $\sigma^{z}_{OH}$ & & $|\sigma_{OH}|$ & & $|\sigma_{SH}|$ & & $E_g (\text{eV})$ & &  $\left[K_1^{(3)}\right]$  & & $\left[K_2^{(3)}\right]$ & & $Q^{(3)}_{\rm c}$ \\ [0.5ex] 
		\hline
		\hline
		MoS$_2$ (42)   & & $0.00$ & & $0.00$ & & $2.65$ & & $2.65$ & & $0.00$ & & $1.60$ & & $-1$ & & $2$ & & $2/3$\\
		\hline
		MoSe$_2$ (42)  & & $0.00$ & & $0.00$ & & $2.63$ & & $2.63$ & & $0.00$ & & $1.34$ & & $-1$ & & $2$ & & $2/3$\\ [0.5ex] 
		\hline
		MoTe$_2$ (42)  & & $0.00$ & & $0.00$ & & $2.47$ & & $2.47$ & & $0.00$ & & $0.95$ & & $-1$ & & $2$ & & $2/3$\\ [0.5ex] 
		\hline
		CrS$_2$ (24) & & $0.00$ & & $0.00$ & & $1.89$ & & $1.89$ & & $0.00$ & & $0.90$ & & $-1$ & & $2$ & & $2/3$\\ [0.5ex] 
		\hline
		CrSe$_2$ (24) & & $0.00$ & & $0.00$ & & $1.92$ & & $1.92$ & & $0.00$ & & $0.71$ & & $-1$ & & $2$ & & $2/3$\\ [0.5ex] 
		\hline
		CrTe$_2$ (24) & & $0.00$ & & $0.00$ & & $1.91$ & & $1.91$ & & $0.00$ & & $0.47$  & & $-1$ & & $2$ & & $2/3$\\ [0.5ex] 
		\hline
		WS$_2$ (74)  & & $0.00$ & & $0.00$ & & $1.43$ & & $1.43$ & & $0.00$ & & $1.56$ & & $-1$ & & $2$ & & $2/3$\\ [0.5ex] 
		\hline
		WSe$_2$ (74)  & & $0.00$ & & $0.00$ & & $1.63$ & & $1.63$ & & $0.00$ & & $1.27$ & & $-1$ & & $2$ & & $2/3$\\ [0.5ex] 
		\hline
		WTe$_2$ (74)  & & $0.00$ & & $0.00$ & & $1.46$ & & $1.46$ & & $0.00$ & & $0.77$ & & $-1$ & & $2$ & & $2/3$\\ [0.5ex] 
		\hline
		
		TiS$_2$ (22)  & & $0.00$ & & $0.00$ & & $0.78$ & & $0.78$ & & $0.00$ & & $0.72$  & & $0$ & & $2$ & & $2/3$\\ [0.5ex] 
		\hline
		TiSe$_2$ (22) & & $0.00$ & & $0.00$ & & $0.75$ & & $0.75$ & & $0.00$ & & $0.53$ & & $0$ & & $2$ & & $2/3$\\  [0.5ex] 
		\hline
	\end{tabular}
\caption{Orbital Hall conductivity of TMD monolayers in the 2H structural phase. In the first column, we list  the compound and the atomic number of its transition metal atom (Z$_{\text{M}}$). Columns 2-5 show the corresponding OHCs inside their insulating gaps ($\sigma^{\eta}_{OH}$, for $\eta=x,y,z$, and $|\sigma_{OH}|$), respectively. All the OHCs are given in units of $e/2\pi$. We note that inside their gaps all the 2H TMD monolayers studied exhibit $\sigma^{x}_{OH}=\sigma^{y}_{OH}=0$. In column 6 we show that 
the spin Hall conductivities $|\sigma_{SH}|$ also vanish, as expected. The band gaps $E_g$ are displayed in column 7. The calculated topological invariants $\left[K_1^{(3)}\right]$ and $\left[K_2^{(3)}\right]$, and the charge of the corner states $Q^{(3)}_{\rm c}$, all defined in the main text, are given in columns 8, 9 and 10, respectively.}
	\label{table_2HTMD_Sup}
\end{table*}

\begin{figure}[!h]
\begin{center}
\includegraphics[width=0.75\textwidth]{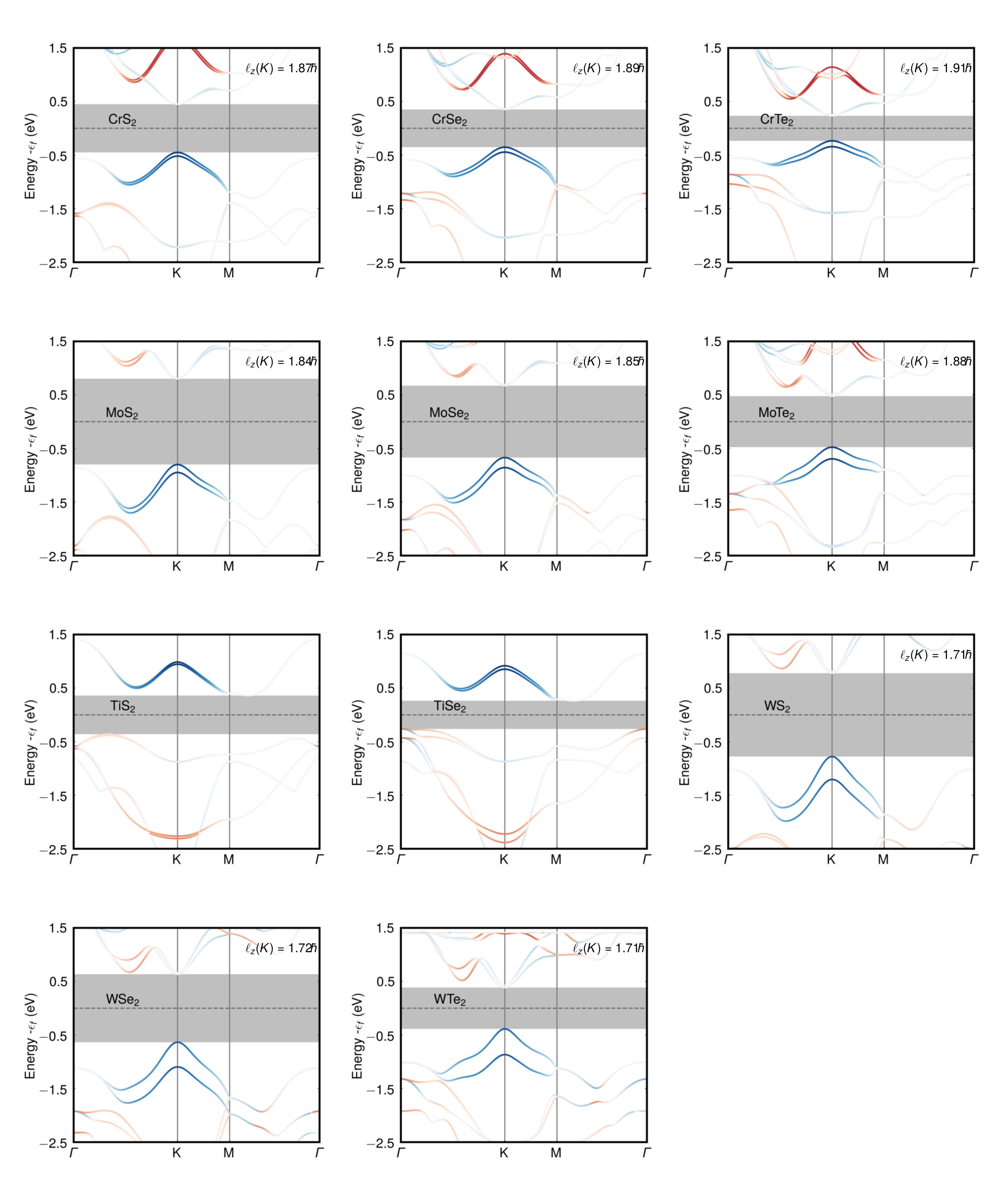}
\end{center}
\caption{Electronic energy bands of the 2H TMD monolayers listed in table \ref{table_2HTMD_Sup}, calculated with the PAO Hamiltonian along some high-symmetry directions of the 2D BZ. The color code illustrates the corresponding expectation values of ${\ell}_z$. For compounds that display direct energy band gap centered at the $K$ and $K'$ symmetry points, we explicitly quote the value of ${\ell}_z(K)$ in the figure. The blue (red) tonalities indicates a polarization of OAM that approaches $2\hbar$ (-$2\hbar$).\label{FigureS1}}
\end{figure}

Fig. \ref{FigureS1} shows the electronic energy bands calculated with the PAO Hamiltonian along some high symmetry directions of the two-dimensional (2D) BZ. The color-code indicates the expectation value of the $\hat{z}$-component of the OAM operator given by
\begin{eqnarray}
 \ell_{z,n}({\bf k})=\langle u_{n,{\bf k}}\big| \hat{L}_z \big| u_{n,{\bf k}}\rangle. 
\label{LZtexture} 
\end{eqnarray}
In our color convention, the blue (red) tonalities represents an OAM polarization that approaches $2\hbar$ (-$2\hbar$), indicating  dominance of the transition-metal atoms' d-shell. We shifted the energy scale to make the origin coincide with the Fermi-level.

\begin{figure}[!h]
\begin{center}
\includegraphics[width=0.75\textwidth]{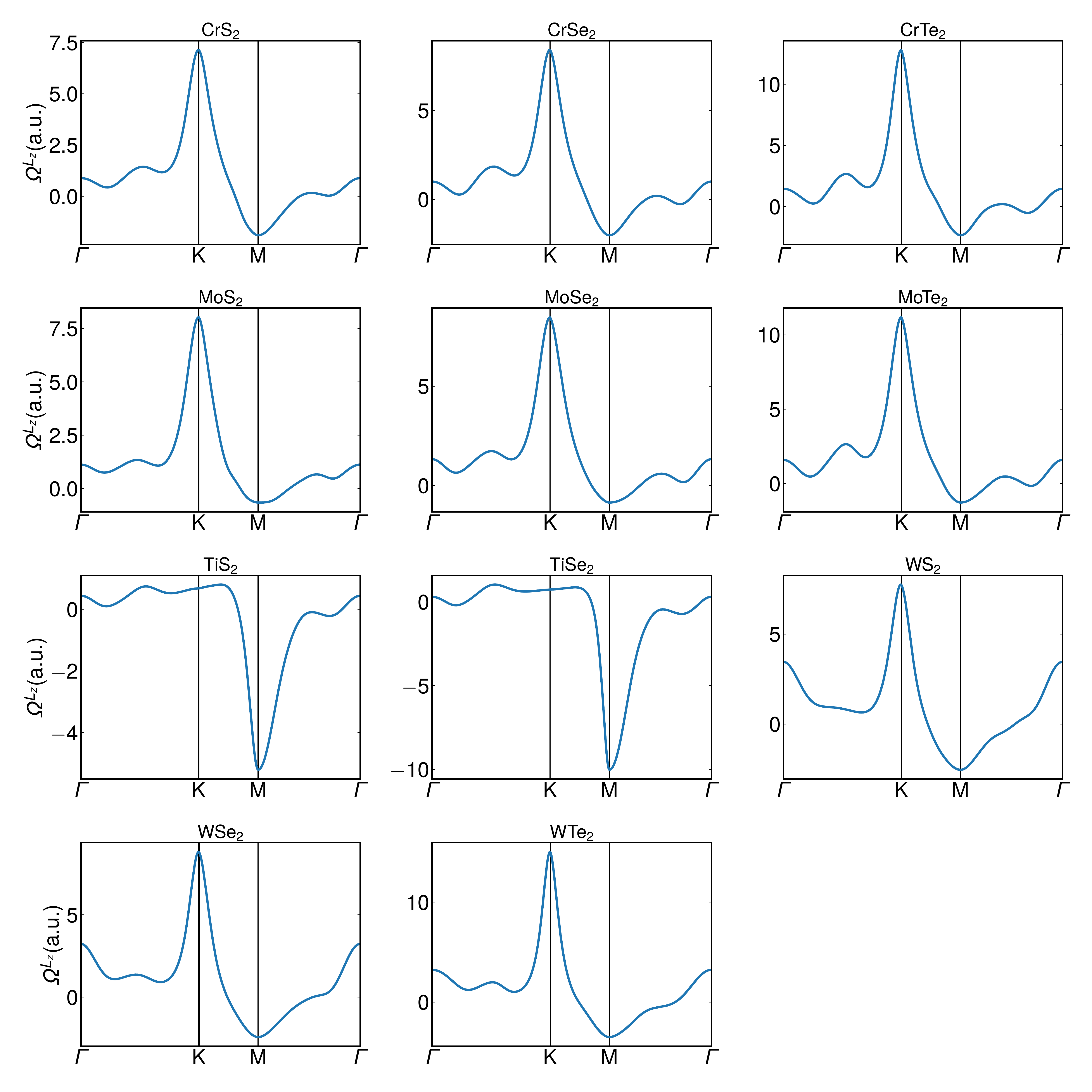}
\end{center}
\caption{Orbital-weighted Berry curvatures of the valence-bands, defined by Eq. (\ref{OWBerryValenceBand}), calculated for all the 2H TMD monolayers listed in Table \ref{table_2HTMD_Sup}, along some high-symmetry directions in the 2D BZ \label{FigureS2}.}
\end{figure}

In Fig. \ref{FigureS2}, we depict the orbital-weighted Berry curvatures of the valence-bands, defined by Eq. (\ref{OWBerryValenceBand}), calculated for all the 2H TMD monolayers listed in the table \ref{table_2HTMD_Sup}, along some high-symmetry directions of the 2D BZ. 

{\it Discussion}
We notice two qualitatively distinct behaviors in the results depicted in Figs. \ref{FigureS1} and \ref{FigureS2}. Mo-, Cr- and W-based systems exhibit a direct enrgy band gap and the top of the valence-band occurs at K valleys, where the eigenstates have strong $d$-character and are highly polarized along the out-of-plane $\hat{z}$-direction, indicated by the intense blue tonality of the energy band spectra in this energy region. The expectation values of the $\hat{z}$ component of the OAM calculated at the top of the valence bands ($\ell_z(K)$) are quoted in the right upper corners of the Fig. \ref{FigureS1} panels. For M=Mo, Cr and W, the orbital-weighted Berry curvatures are strongly peaked around the K-valleys. This means that the physics of OAM can be well described by a Dirac model \cite{Us3} whose valence-band involves the linear combinations of transition metal d states $\big( \big|d_{x^2-y^2}\rangle-i\tau \big|d_{xy}\rangle \big)/\sqrt{2}$  which carry OAM $2\tau \hbar$, where $\tau=\pm 1$ for K and K' valleys. However, for Ti , the transition-metals d-shell-band is positioned above the electronic band-gap, i.e., in the conduction band. On the other hand, the gap is indirect, and $\Omega^{L_z} ({\bf k})$ is more spread throughout the BZ. Consequently, the OHC plateau in the insulating gap $|\sigma_{OH}|$ for H-TMDs of Ti is always smaller than $1.0 \times (e/2\pi)$ [see table \ref{table_2HTMD_Sup}].

\section{1T-TMDs}

\begin{table*}[h!] 
\vspace{+1em}

	\centering
	\begin{tabular}{||c c || c c | c c | c c | c c | c c | c c | c ||} 
		\hline
		1T-TMD (Z$_\text{M}$)& & $\sigma^{x}_{OH}$ & &  $\sigma^{y}_{OH}$ & &  $\sigma^{z}_{OH}$ & & $|\sigma_{OH}|$ & & $|\sigma_{SH}|$ & & $E_g (\text{eV})$ & & $\mathbb{Z}_4$    \\ [0.5ex] 
		\hline
		\hline
		NiS$_2$ (28) & & $0$ & & $1.02$ & & $-1.78$ & & $2.05$ & & $0.00$ & & $0.54$ & & 2   \\
		
		\hline
	        PdS$_2$ (46)  & & $0$ & & $0.73$ & & $-1.22$ & & $1.42$ & & 0.00 &&$1.14$ && 2  \\
		
		\hline
		PdSe$_2$ (46) & & $0$ & & $0.85$ & & $-1.65$ & & $1.85$ & & $0.00$ & & $0.52$ & & 2  \\
		\hline
		PtTe$_2$ (78) & & $0$ & & $0.82$ & & $-1.59$ & & $1.70$ & & $0.00$ & & $0.37$ & & 2  \\
		\hline
		PtSe$_2$ (78)  & & $0$ & & $0.60$ & & $-1.01$ & & $1.18$ & & $0.00$ & & $1.17$ & & 2  \\
		\hline
	    PtS$_2$ (78) & & $0$ & & $0.51$ & & $-0.77$ & & $0.92$ & & $0.00$ & & $1.72$ & & 2  \\
		\hline
		ZrS$_2$ (40) & & $0$ & & $0.61$ & & $-0.34$ & & $0.70$ & & $0.00$ & & $1.16$ & & 2  \\
		\hline
		ZrSe$_2$ (40) & & $0$ & & $0.84$ & & $-0.58$ & & $1.02$ & & $0.00$ & & $0.34$ & & 2  \\
		\hline
		HfS$_2$ (72)  & & $0$ & & $0.32$ & & $-0.08$ & & $0.33$ & & $0.00$ & & $1.27$ & & 2  \\
		\hline
		HfSe$_2$ (72)  & & $0$ & & $0.71$ & & $-0.40$ & & $1.02$ & & $0.00$ & & $0.48$ & & 2  \\
		\hline
	
		NiO$_2$ (28)  & & $0$ & & $0.51$ & & $-0.76$ & & $0.91$ & & $0.00$ & & 1.25 & & 2  \\
		\hline
		PdO$_2$ (46) & &	$0$& &	0.37	& &	-0.46	& &	0.59	& &	0.00	& &	$1.39$ & & 2 \\
		\hline
		PtO$_2$ (78) & &	0	& &	0.26	& &	-0.27& &	0.37	& &	0.00	& &	$1.70$ & & 2 \\
		\hline
		
	\end{tabular}
	
\caption{Orbital Hall conductivity of TMD monolayers in the 1T structural phase. In the first column, we list  the compound and the atomic number of its transition metal atom (Z$_{\text{M}}$). Columns 2-5 show the corresponding OHCs within their insulating gaps ($\sigma^{\eta}_{OH}$, for $\eta=x,y,z$, and $|\sigma_{OH}|$), respectively. All the OHCs are given in units of $e/2\pi$. In column 6 we show that 
the spin Hall conductivities $|\sigma_{SH}|$ vanish, as expected. The band gaps $E_g$ are  displayed in column 7, and column 8 shows that the topological indicator $\mathbb{Z}_4 = 2$, for all elements in this table.}
	\label{table_1TTMD_Sup}
\end{table*}

Table \ref{table_1TTMD_Sup} extends the Table II presented in the main text. It includes all insulating TMD monolayers that naturally stabilize in the 1T structural phase. They are listed here in decreasing order of their orbital Hall conductivity plateaus.

\begin{figure}[!h]
\begin{center}
\includegraphics[width=0.99\textwidth]{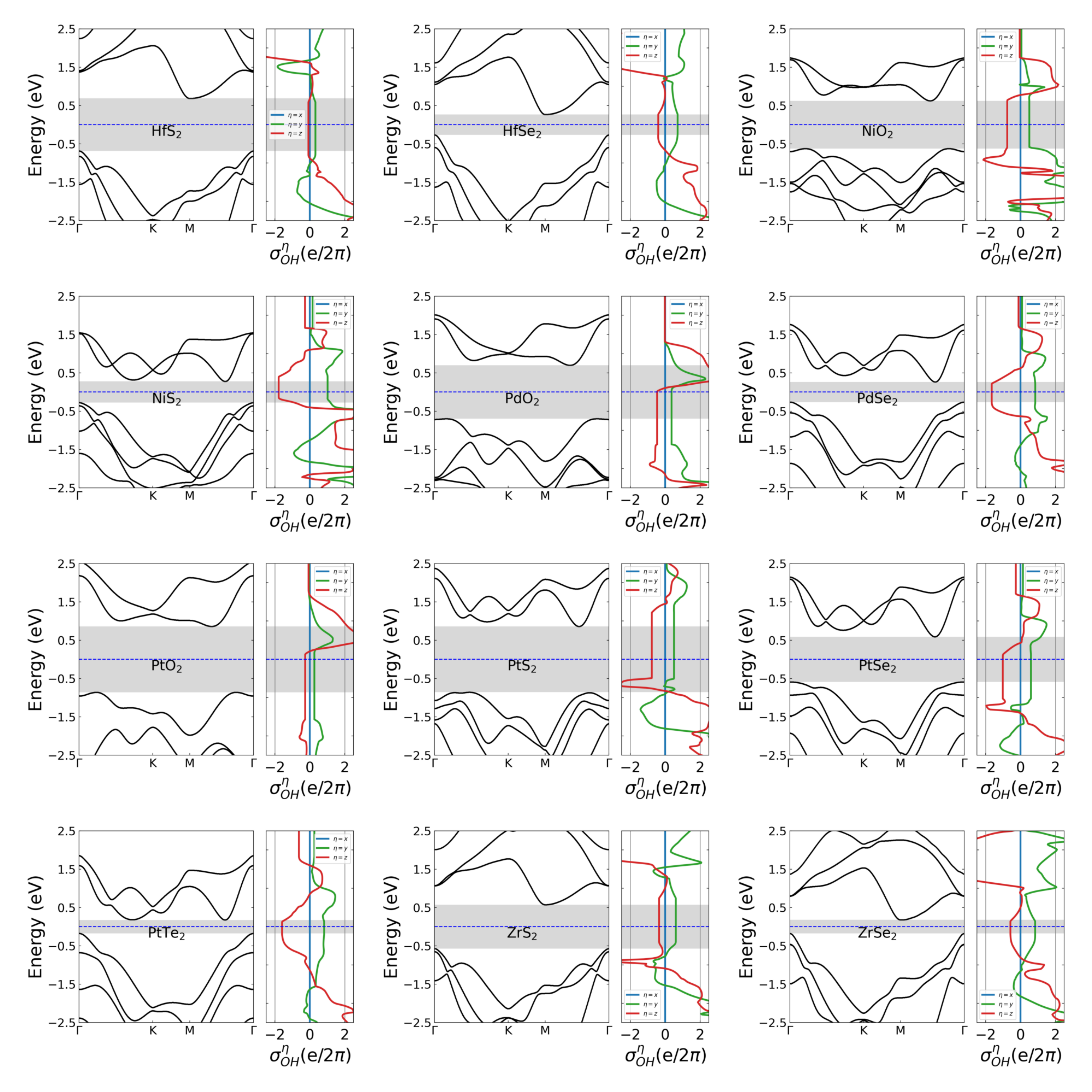}
\end{center}
\caption{Electronic energy bands calculated along some high-symmetry directions of the 2D BZ are exhibited in the left panels for all 1T TMD monolayers listed in Table \ref{table_1TTMD_Sup}. The corresponding orbital Hall conductivities $\sigma^{x}_{OH}$, $\sigma^{y}_{OH}$ and $\sigma^{z}_{OH}$, calculated as functions of energy, are depicted by the green, blue and red lines, respectively,in the right panels.\label{FigureS3}}
\end{figure}
{\it Discussion}
TMD monolayers in the 1T structural phase are centrosymmetric. As a result, the band-structure is spin degenerate and the expectation value $\langle \ell_z\rangle_{n, {\bf k}}=0$ for the spin degenerate bands. 

Fig. \ref{FigureS3} shows the electronic energy bands calculated along some high-symmetry directions of the 2D BZ, snd the orbital Hall conductivities $\sigma^{x}_{OH}$, $\sigma^{y}_{OH}$ and $\sigma^{z}_{OH}$, calculated as functions of energy, for all 1T TMD monolayers listed in table \ref{table_1TTMD_Sup}. Differently from the 2H structures, most 1T TMDs have indirect energy band gaps. Also, the top of the valence bands have contribution of different sets of states that can behave as pseudo-spinors. This results in orbital-weighted Berry curvatures that have two different components and are spread over the whole Brillouin zone. 

The orbital Hall conductivity also has two different components, although $L_z$ is still the main contribution to the orbital Hall effect in the insulating phase.

\begin{figure}[!h]
\begin{center}
\includegraphics[width=0.75\textwidth]{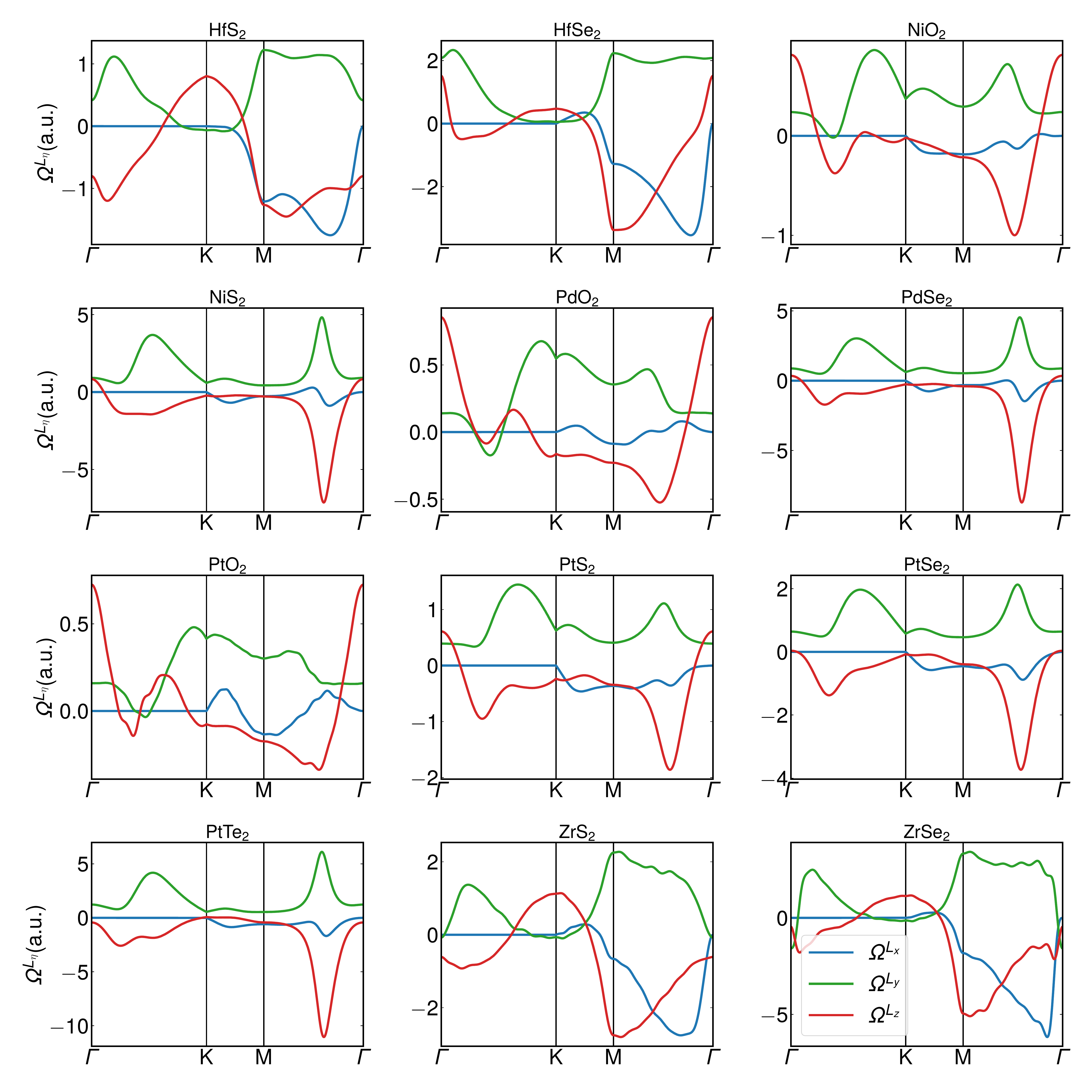}
\end{center}
\caption{For each 1T monolayer TMD presented in the table, we show the orbital-weighted Berry-curvature of the valence-band for the three components of the OAM.\label{FigureS4} }
\end{figure}

\section{Connection between higher order topological insulators and orbital Hall effect in a low energy model}

{
To show a direct connection between HOTIs and the orbital Hall effect, let us consider the effective model presented in Ref. \onlinecite{Schindler2018} to characterize the higher-order topological phase of Bismuth. In that case, the topology of Bismuth is protected by the combination of the $\hat{C}_3$ rotation, $\mathcal{I}$ and $\mathcal{T}$ symmetries, and the authors demonstrated that the existence of multiple band inversions occurring within subspaces characterized by their rotation eigenvalues is overlooked by the first-topological index proposed by Fu and Kane \cite{FuKaneIndexFirstOrder}. Similarly to Bismuth, the two classes of TMDs discussed here (1T and 1H) also present two band inversions. To construct an effective low-energy model for Bismuth, the authors used a Dirac model representation of a HOTI based on the Bernevig-Hughes-Zhang (BHZ) model for topological insulators. It consists of two diagonal blocks containing two BHZ-like models that respect $\hat{C}_3$ rotation symmetry and $\mathcal{I}$. 

Individually, they have a single band inversion and are topological insulators. When connected, they present two band inversions and are indexed as HOTI. Let us start exploring the characteristics of a single BHZ Hamiltonian. Following the supplementary material of Ref. \onlinecite{Schindler2018}, if one considers the basis $\{
|p_+\uparrow\rangle, 
|d_+\downarrow\rangle,
|p_-\downarrow\rangle, 
|d_-\uparrow\rangle
\}
$ where $p_\pm=p_x\pm\mathrm{i}p_y$ and $d_\pm=d_{xy}\pm\mathrm{i}d_{x^2-y^2}$, the BHZ model can be written as:
}

\begin{equation}
\label{HT}
{\cal{H}}_{T} (\mathbf{k})=
\begin{bmatrix} 
\cal{H}(\mathbf{k}) &0 \\
0 &\cal{H}^*(-\mathbf{k})
\end{bmatrix}
\end{equation}
where 
\begin{equation}
\cal{H}(\mathbf{k}) =\begin{bmatrix}
M(k) & v_f k_+\\
  v_fK_- &  -M(k)\\
\end{bmatrix},
\end{equation}
where $k^2=k_x^2+k_y^2$,  $k_\pm=k_x\pm\mathrm{i}k_y$. The effective mass $M(\mathbf{k})$ is given by $M(\mathbf{k})=B-M k^2$, where $B$ and $M$ are two real constants. The transition between the topological and trivial insulating phases occurs when $M$ changes sign.

The Berry curvature of the valence band of each block is given by
$$\Omega^s(k)=s\frac{v_f^2(M+Bk^2)}{((M-Bk^2)^2+v_f^2k^2)^{3/2}}$$
where $s=\pm$ is the block index, $v_f$ represents the Fermi velocity, and

\begin{figure}[!h]
    \centering
    \includegraphics[width=0.4\linewidth]{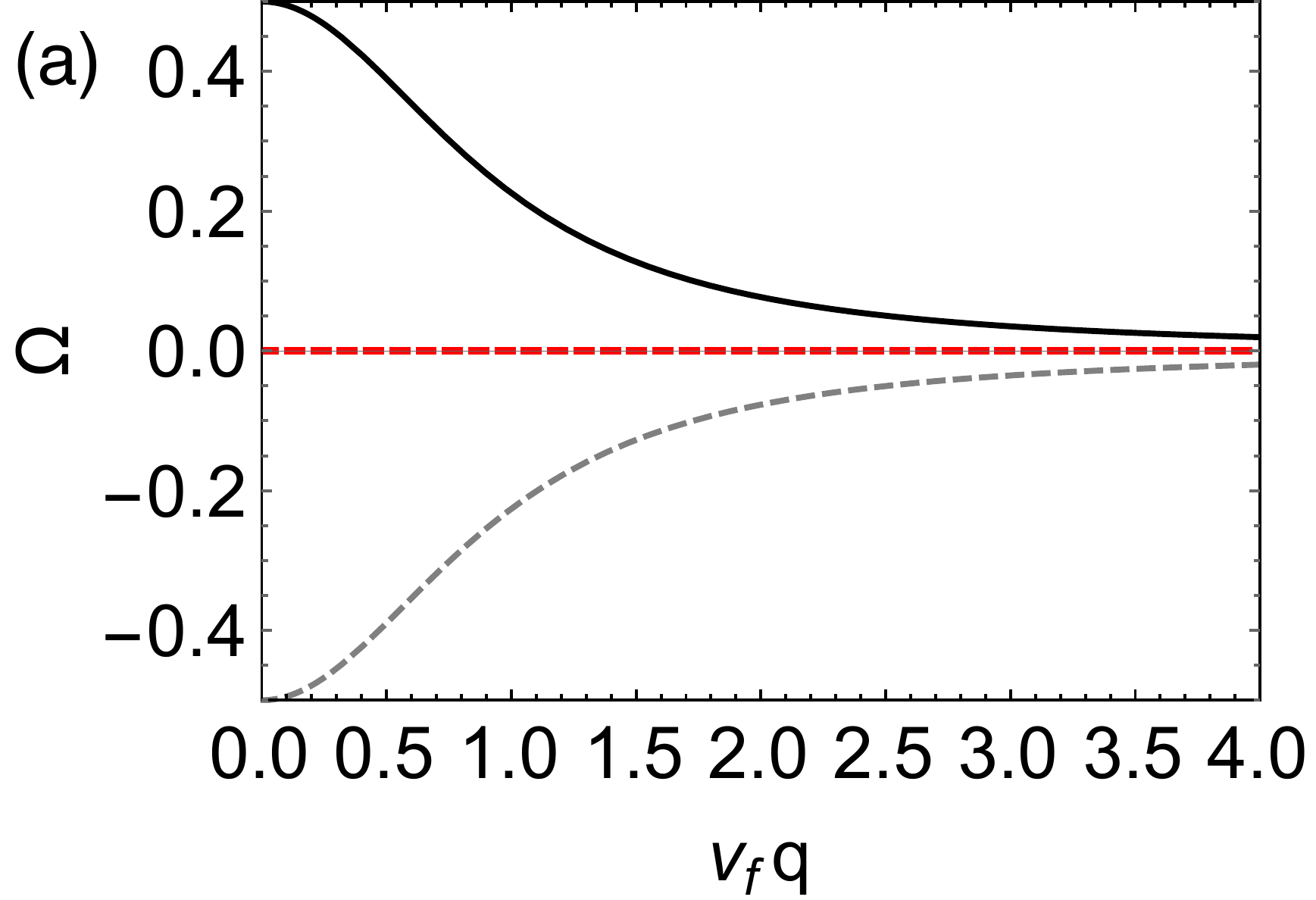}
    \includegraphics[width=0.4\linewidth]{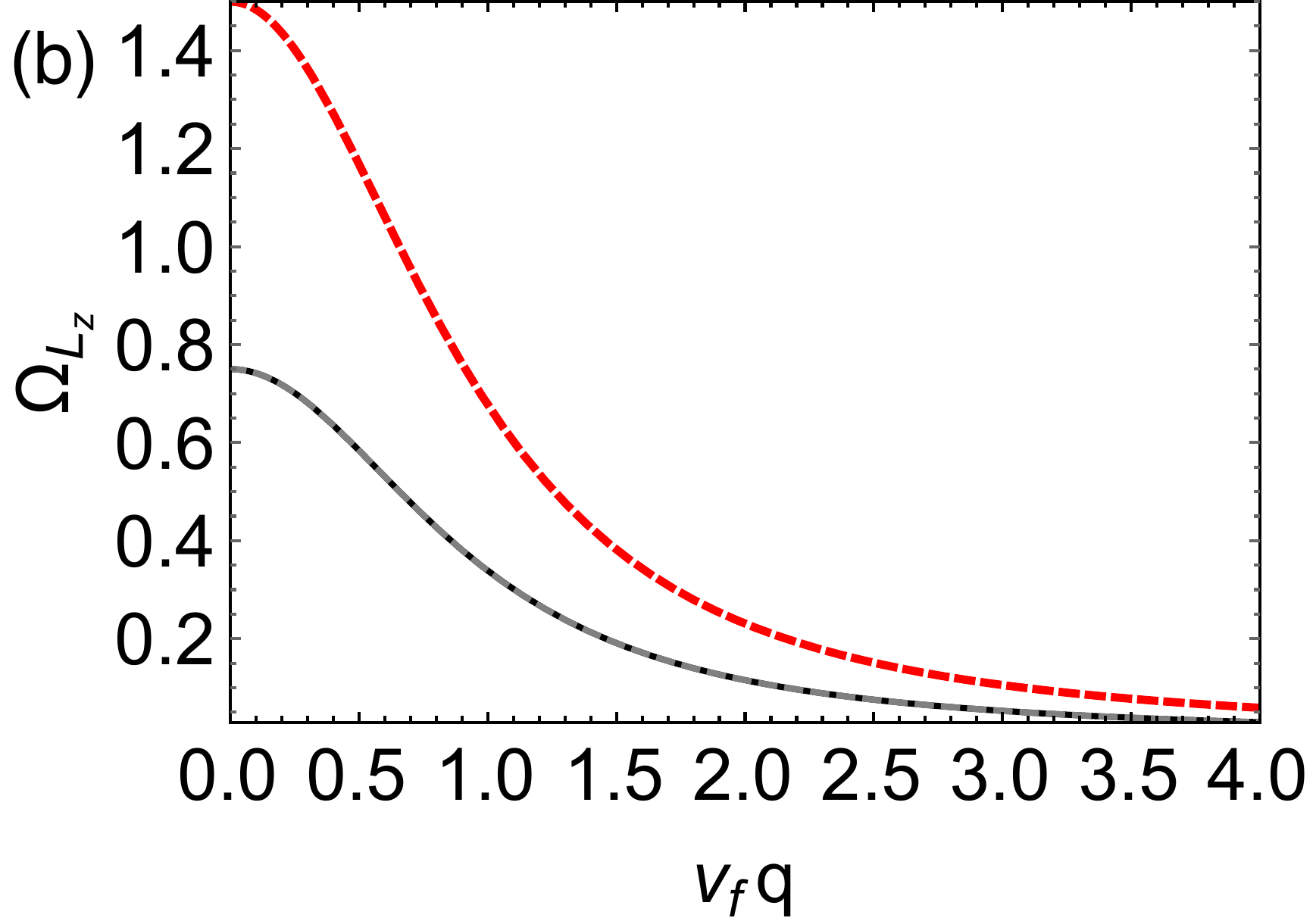}
    \caption{Berry (left panel) and orbital-weighted Berry curvatures (right panel) calculated for the BHZ model with $M=1$ and $B=0.1$. The black solid and grey dashed lines represent the contributions from the $s=+$ and $s=-$ blocks, respectively. Note that the two contributions are identical in panel (b). The red dashed lines depict the sum of the $s=\pm$ contributions that define the corresponding Berry and orbital-weighted Berry curvatures.}
    \label{fig:BerryH1}
\end{figure}

{ The $s=\pm$ contributions of each block to the total Berry curvature are depicted in the left panel of figure \ref{fig:BerryH1}.  They exhibit an extremum at $k$=0, where the band inversion occurs, and tend to zero as $\sim \pm 1/k$ for large values of $k$. We note that even for $B=0$, the system is a $\mathbb{Z}_2$ topological insulator, and the total Berry curvature for this model is zero, as expected for time-reversal symmetric systems. In contrast, the $s=\pm$ contributions from each block to the orbital-weighted Berry curvature are identical and given by}

$$\Omega^L(k)=\frac{3\hbar}{4}\frac{v_f^2(M+Bk^2)}{((M-Bk^2)^2+v_f^2k^2)^{3/2}}.$$

{
They are portrayed in the right panel of figure \ref{fig:BerryH1}  and clearly lead to a non-null total orbital Berry curvature, whose $\sim 1/k$ decay for large $k$ implies that the BHZ model describes a $\mathbb{Z}_2$ topological insulator with a sizable orbital Hall effect.\\}

{
The Dirac representation of a HOTI with $\mathcal{T}$,$\hat{C}_3$ and $\mathcal{I}$ symmetries in the basis of orbitals $\{
|p_-\uparrow\rangle, 
|d_-\downarrow\rangle,
|p_+\downarrow\rangle, 
|d_+\uparrow\rangle,
|p_+\uparrow\rangle, 
|d_+\downarrow\rangle,
|p_-\downarrow\rangle, 
|d_-\uparrow\rangle
\}$ can be expressed as 
\begin{equation}
H(\mathbf{k}) = H_1(\mathbf{k}) \oplus H_3(\mathbf{k}),
\label{eq: continuum H}
\end{equation}
where $H_1(\mathbf{k})$ and $H_3(\mathbf{k})$ are disconnected blocks of the Hamiltonian characterized by the rotation eigenvalues $-1$ and $e^{\pm i\pi/3}$, respectively. Each block is defined by 

\begin{equation}
\label{H_j}
H_j(\mathbf{k}) =
\begin{pmatrix} M((\mathbf{k})) &  v_f k_+^j & 0 &  0 \\ 
 v_f k_-^j & -M(\mathbf{k}) &  0 & 0 \\
0 &  0 & M(\mathbf{k}) & - v_f k_-^j \\
 0 & 0 & - v_f k_+^j & - M(\mathbf{k})
\end{pmatrix},
\end{equation}
for $j=1,3$. Here $k_\pm=k_x\pm\mathrm{i}k_y$, $k_{\pm}^j= (k_{\pm})^j$,.  We note that $H_1(\mathbf{k})$ represents the BHZ model given by Eq. \ref{HT}. The off-diagonal terms in each $H_3(\mathbf{k})$ block are altered to $k_{\pm}^3$ in order to comply with $\mathcal{I}$ and the corresponding representation of $\hat{C}_3$. Nevertheless, $H_3(\mathbf{k})$ also describes a $\mathbb{Z}_2$ TI.

The Berry curvature of the valence band associated with each sub-block of $H_3(\mathbf{k})$ is given (in polar coordinates) by
$$\Omega^s(k)=s\frac{3}{4}\frac{k^4(3M-Bk^2)}{((M-Bk^2)^2+v_f^2k^6)^{3/2}}$$
where $s=\pm$ is the sub-block index.

\begin{figure}[!h]
    \centering
    \includegraphics[width=0.4\linewidth]{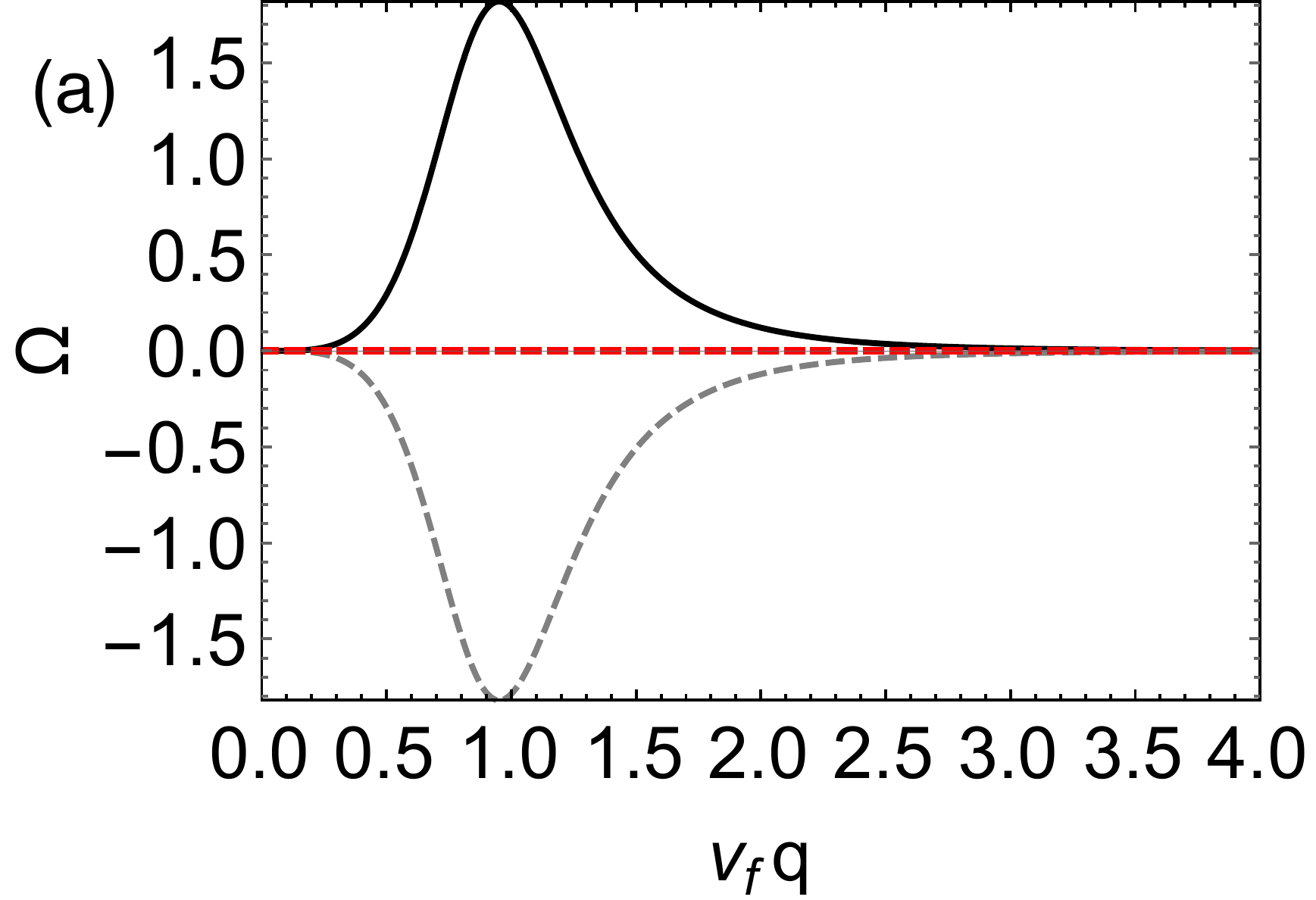}
    \includegraphics[width=0.4\linewidth]{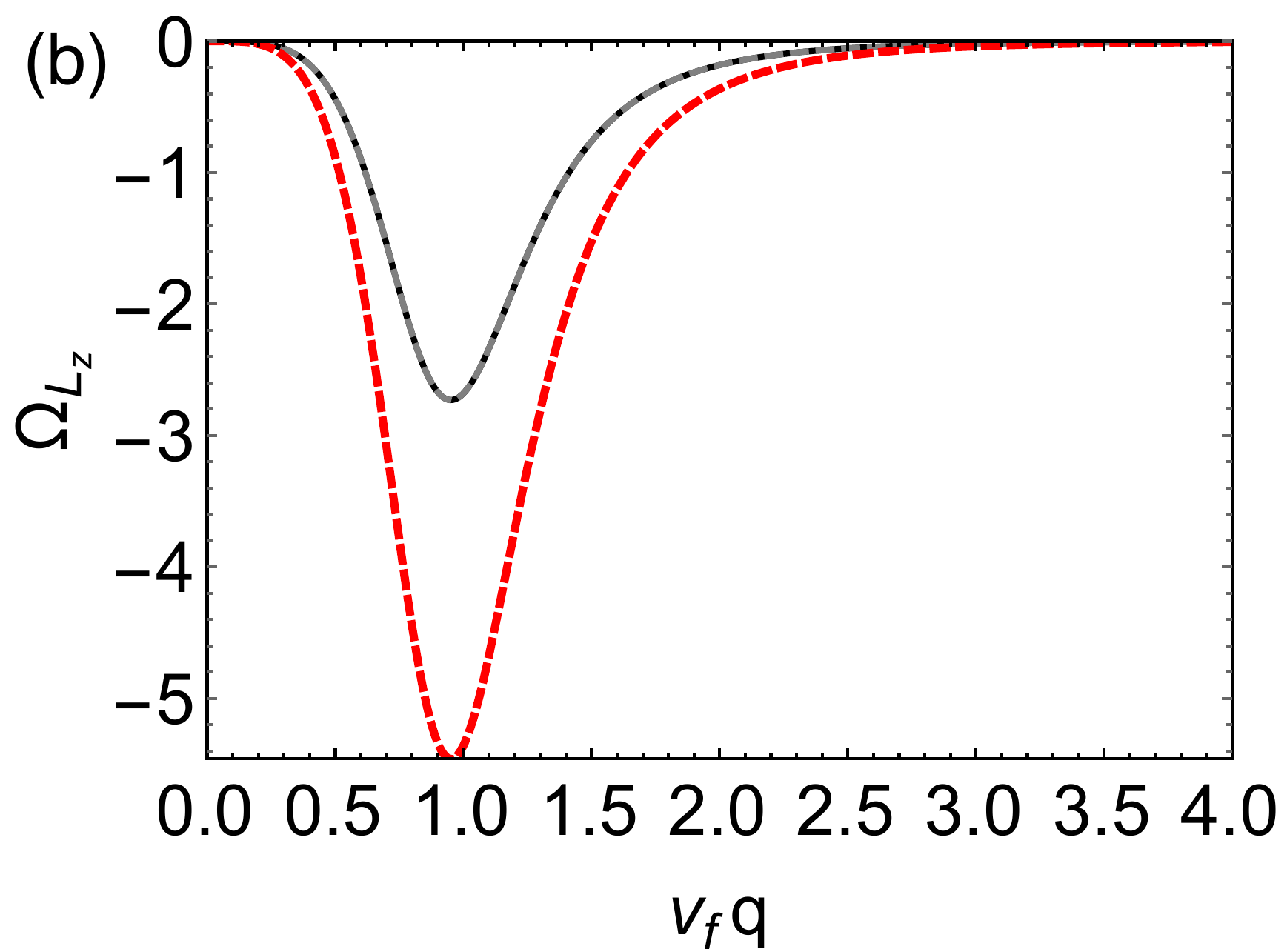}
    \caption{Berry (left panel) and orbital-weighted  Berry curvatures (right panel) calculated for the $H_3$ model with $M=1$ and $B=0.1$. The black solid and grey dashed lines correspond to the contributions from the $s=+$ and $s=-$ blocks, respectively. The red dashed lines depict the total Berry and orbital-weighted Berry curvatures, which are given by the corresponding sums of the $s=+$ and $s=-$ contributions.}
    \label{fig:BerryH3}
\end{figure}

In the left panel of figure \ref{fig:BerryH3}, we note that due to the modification of the off-diagonal terms in $H_3(\mathbf{k})$ the extrema of $s=\pm$ contributions to the Berry curvature are located at $k\ne0$, and for large values of $k$ they decay as $\sim \pm 1/k^3$. Similarly to $H_1(\mathbf{k})$, even for $B=0$, the system is a $\mathbb{Z}_2$ topological insulator, and the total Berry curvature for the model vanishes, as expected for time-reversal symmetric systems. However, the $s=\pm$ contributions to the orbital-weighted Berry curvature are identical and given by  
$$\Omega^L(k)=-\frac{9\hbar}{4}\frac{v_f^2k^4(M-Bk^2)}{((M-Bk^2)^2+v_f^2k^6)^{3/2}}.$$
The right panel of figure \ref{fig:BerryH3}  depicts the $s=\pm$ contributions and the total orbital-weighted Berry curvature for the $H_3(\mathbf{k})$ model. The non-vanishing total orbital Berry curvature and its $1/k^3$ decay for large $k$ imply that it also describes a $\mathbb{Z}_2$ topological insulator with a sizable orbital Hall effect. Since $H_1(\mathbf{k})$ and $H_3(\mathbf{k})$ present a band inversion, the complete Hamiltonian represents a HOTI~\cite{Schindler2018} with finite orbital Hall effect as the total orbital Berry curvature remains sizable. This explicitly shows that any HOTI that is reasonably described by the model Hamiltonian given by Eqs. \ref{eq: continuum H} and \ref{H_j} will exhibit finite orbital Hall conductivity.   

}

\section{Connection between higher-order topological insulators and orbital Hall effect in a multi-orbital triangular lattice for 2H-TMDs}

{ Here, we follow a recipe to obtain an electronic 2D HOTI on a multiorbital triangular lattice, as presented in reference \onlinecite{Eck2022}. One key aspect of this approach is the possibility of tuning the appearance of different topological phases by changing a few parameters. In this work, Eck \textit{et al.} demonstrated that the essential ingredients for the emergence of the HOTI phase are the spin-orbit coupling and two terms that break the inversion and the horizontal mirror symmetries. Surprisingly, the three-band model commonly used to describe the low-energy electronic properties of 2H TMDs falls within the same class of triangular lattice systems. To illustrate this, we will focus on the competition between the spin-orbit coupling and terms that break the inversion symmetry of the Hamiltonian.
Using the basis $\{d_{z^{2}},d_{xy},d_{x^{2}-y^{2}}\}$,the three bands model can be written as

\begin{equation}
H=\begin{bmatrix}h_{0} & h_{1} & h_{2}\\
h_{1}^{*} & h_{11} & h_{12}\\
h_{2}^{*} & h_{12}^{*} & h_{22}
\end{bmatrix},
\label{eqn:Hamiltonian3Bands}
\end{equation}
where
\begin{align}
&h_{0}=2t_{0}(\cos2\alpha+2\cos\alpha\cos\beta)+\epsilon_{1}, \\
&h_{1}=-2\sqrt{3}t_{2}\sin\alpha\sin\beta+2it_{1}(\sin2\alpha+\sin\alpha\cos\beta)\\
&h_{2}=2t_{2}(\cos2\alpha-\cos\alpha\cos\beta)+2\sqrt{3}it_{1}\cos\alpha\sin\beta, \\
&h_{11}=2t_{11}\cos2\alpha+(t_{11}+3t_{22})\cos\alpha\cos\beta+\epsilon_{2},\\
&h_{22}=2t_{22}\cos2\alpha+(3t_{11}+t_{22})\cos\alpha\cos\beta+\epsilon_{2},\label{eq:H2222}\\
&h_{12}=\sqrt{3}(t_{22}-t_{11})\sin\alpha\sin\beta
+4it_{12}\sin\alpha(\cos\alpha-\cos\beta)\\
&\alpha=\frac{1}{2}k_{x}a \mbox{ and } \beta=\frac{\sqrt{3}}{2}k_{y}a.
\end{align}
The spin-orbit coupling can be written as
\begin{equation}
H_{\rm SOC}=s_z\frac{\lambda}{2}\begin{bmatrix}0 & 0 & 0\\
0 & 0 & 2i\\
0 & -2i & 0
\end{bmatrix},
\end{equation}

where $\lambda$ is the strength of the coupling and $s_z=\pm 1/2$ are the eigenstates of the z component of the spin operator.\\ 

From equation \eqref{eqn:Hamiltonian3Bands},  we identify that the terms proportional to $t_1$, $t_2$, and $t_{12}$ are responsible for the inversion symmetry breaking of the system. To illustrate the transition from a HOTI with sizable orbital Hall conductivity to a trivial insulating phase, we will consider the tight-binding parameters of MoS${}_2$ as described in Ref. \onlinecite{ThreeBandTMD} but increasing the SOC to $\lambda_{SOC}=15\times\lambda_{\text{MoS}{}_2}$. We focus on two situations. In the first case, we consider the system to have broken inversion symmetry, whereas in the second case, we restore the inversion symmetry of the system and neglect the hybridizations between orbitals in different irreducible representations by setting $t_1=t_2=t_{12}=0 $.\\

\begin{figure}[!h]
    \includegraphics[width=0.6\linewidth]{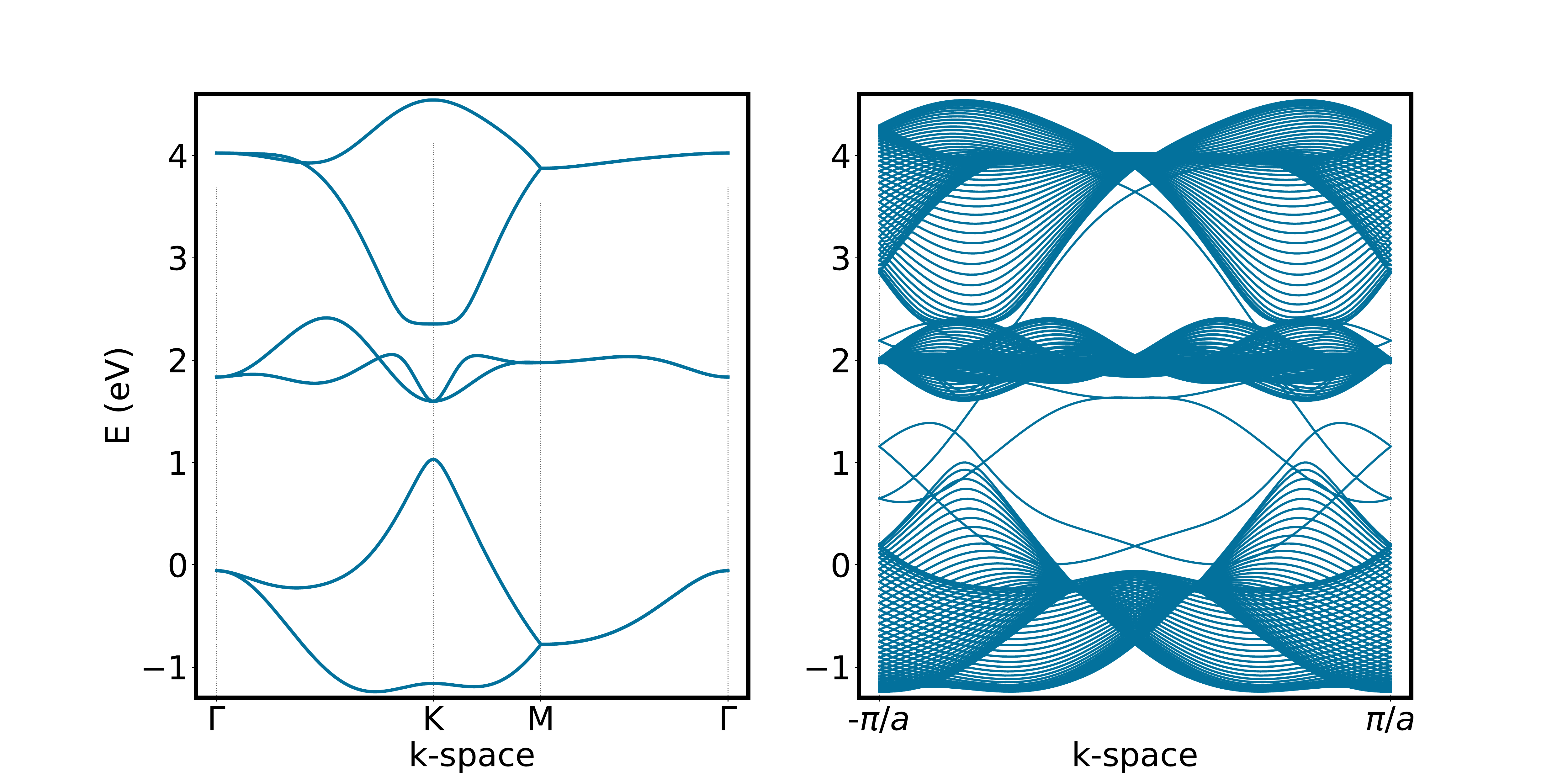}
    \includegraphics[width=0.6\linewidth]{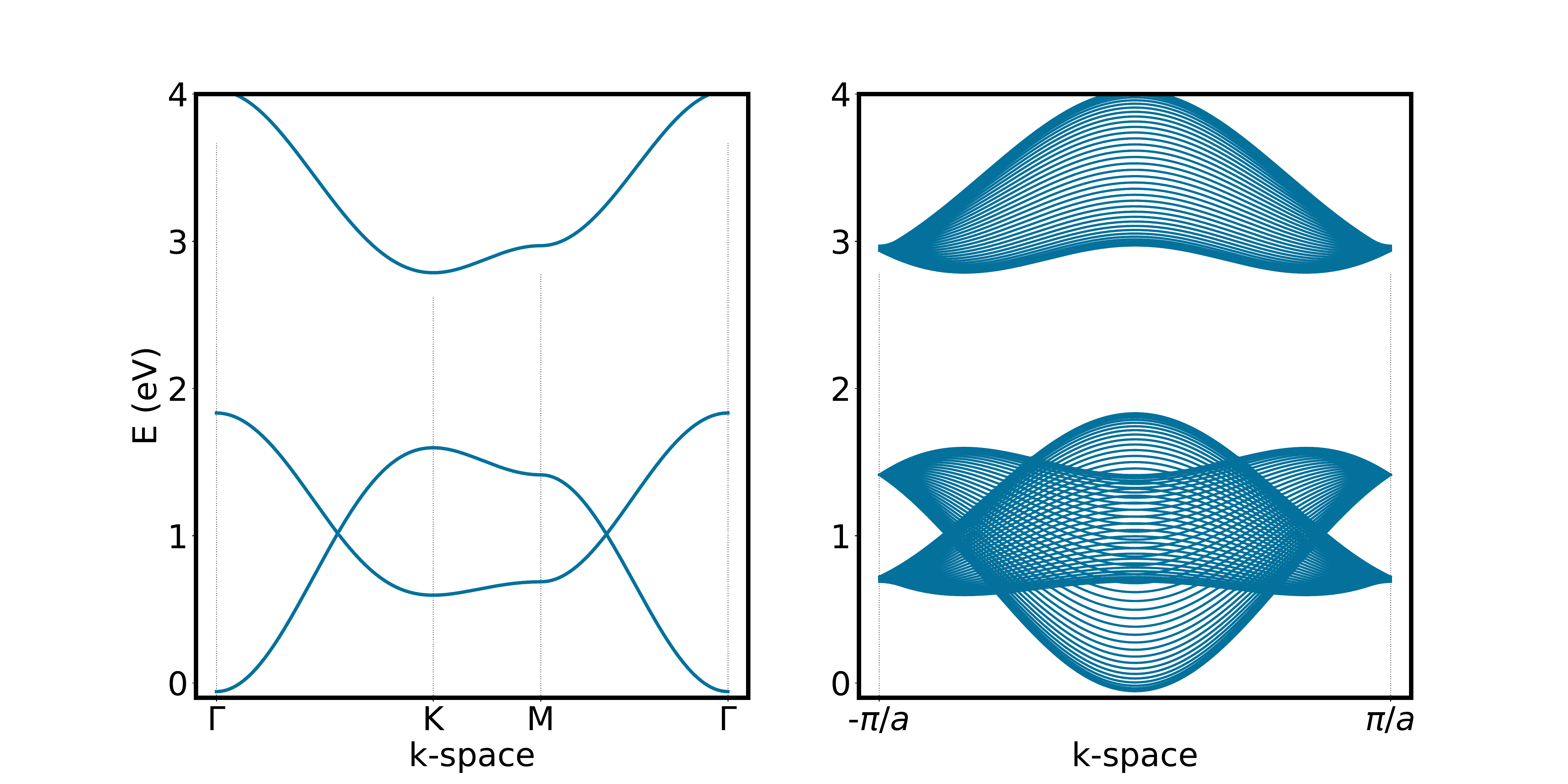}
    \caption{Upper panel: bulk (left) and nanoribbon (right) band structures for the three bands model of MoS$_2$ with strong SOC (case A). Lower panel:  bulk (left) and nanoribbon (right) band structures for the three bands model of MoS$_2$ with strong SOC and $t_1=t_2=t_{12}=0$ (case B). }
    \label{fig:strongsoc}
\end{figure}
\begin{figure}[!h]
    \includegraphics[width=0.4\linewidth]{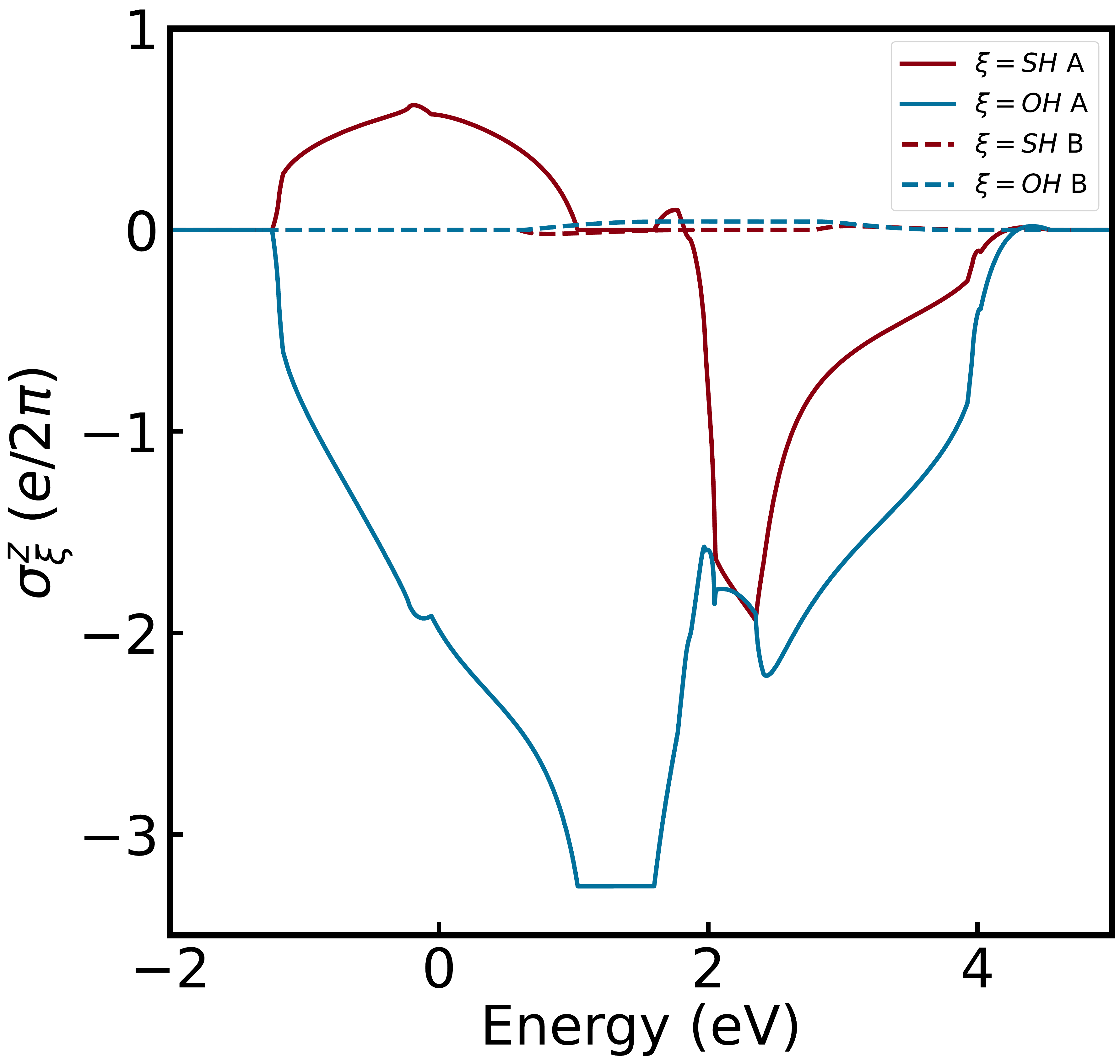}
\caption{Orbital Hall conductivity (blue line) and spin Hall conductivity (red) for the three bands model of MoS$_2$ with strong SOC (case A - solid) and for the three bands model of MoS$_2$ with strong SOC and $t_1=t_2=t_{12}=0$ (case B - dashed line).}
 \label{fig:OHE}
\end{figure}

Figure \ref{fig:strongsoc} shows the bandstructures of the bulk systems and the bandstructures of nanoribbons extracted from them, whereas figure \ref{fig:OHE} presents the orbital Hall conductivity for the two cases. In the first case (A), we found a strong splitting of the conduction band due to the increased SOC strength. However, the number of edge states within the energy gap separating the valence and conduction bands remains finite. Additionally, it is noticeable that these states do not connect the conduction and valence bands, as in the case of MoS${}_2$. We can see that the orbital Hall conductivity plateau has a height comparable to the one for MoS$_2$ with weak SOC and the spin Hall conductivity goes to zero within the energy gap confirming that despite the increment in $\lambda_{SOC}$ the system remains within the orbital Hall insulating phase. If we turn off the couplings that break inversion symmetry and the couplings that hybridize different irreducible representations ($t_1=t_2=t_{12}=0$), we can see that in the second case (B) the strong SOC is responsible for opening a band gap. However, different from the previous case, the bandstructure of its nanoribbon does not have in-gap edge states, and the system is a trivial insulator with vanishing orbital Hall conductivity.\\

The different orbital properties in the two cases previously discussed can be understood in terms of the orbital hybridizations in the system. In the first case, the presence of $t_1$,$t_2$ and $t_{12}$ not only breaks the inversion symmetry but allows the hybridization between the $d_{z^2}$ orbitals and the orbitals $d_{xy}$ and $d_{x^2 -y^2}$, whose superposition can carry non-zero orbital angular momentum. Thus allowing the appearance of orbital textures on reciprocal space and leading to the sizable orbital Berry curvature manifested in the SOC-independent orbital Hall conductivity plateau displayed in figure \ref{fig:OHE}. Nonetheless, in the second case, the opposite occurs. Restoring the inversion symmetry also increases the rotational symmetry of the system, which changes the point group symmetry of the system from $D_{3h}$ to $D_{6h}$ as a consequence of inhibiting hybridizations between the orbitals of the system. Thus there are no orbital angular momentum textures and the formation of orbital pseudospinors in the system is heavily reduced, and now it is a trivial insulator dominated by the SOC. Our analysis and the appearance of a HOTI phase when lattice symmetries are broken and produce band inversions are consistent with the results from Ref. \onlinecite{Eck2022}.
}

\section{Orbital character of PtS${}_2$ bands}

\begin{figure}[!h]
\begin{center}
\includegraphics[width=0.75\textwidth]{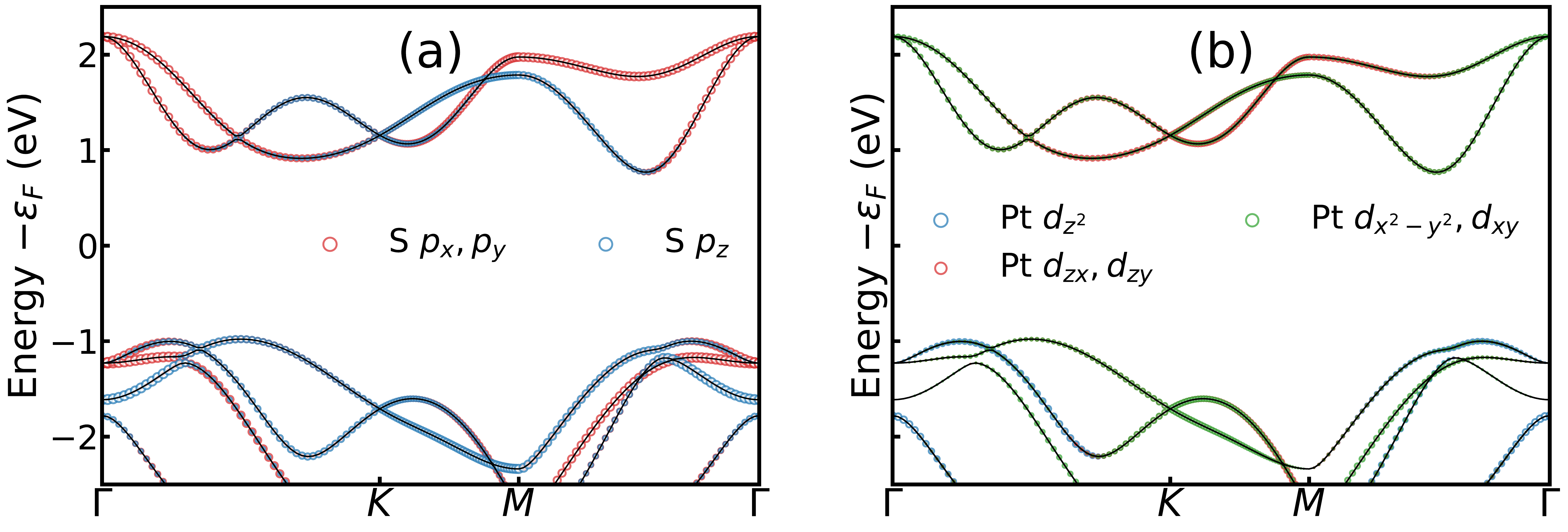}
\end{center}
\caption{Orbital projection of the bulk energy bands of PtS${}_2$.
(a) Contribution from the $p_x,p_y$ (red circles) and $p_z$ (blue circles) orbitals of the Sulfur atoms. (b) Contribution from the $d_{z^2}$ (blue circles), $d_{xz},d_{yz}$ (red circles) and $d_{xy},d_{x^2 - y^2}$ (green circles) orbitals of the Platinum atoms.\label{FigureS5} }
\end{figure}

As mentioned in the main text, the $D_{3d}$ symmetry of the 1T transition metal dichalcogenides imposes that the $p$ orbitals of the sulfur atoms will significantly contribute to upper valence energy bands near $\Gamma$. Figure \ref{FigureS5} portrays the orbital projections of the energy bands for PtSe${}_2$. Panel (a) shows the contribution to the energy states from the $p$ orbitals of the sulfur atoms, and panel (b) corresponds to the $d$ orbitals of the transition metal. From the orbital projections highlighted in panel (a), it is clear that the $p_x$ and $p_y$ orbitals have the dominant contribution to the energy states of the conduction band around $\Gamma$. This state arrangement occurs due to the local asymmetry experienced by the sulfurs on each atomic plane, as we mentioned in the main text. However, as panel (b) portrays, afar from $\Gamma$, the dominant character for the energy states will be given by the $p_z$ orbitals and the pseudospinors formed by the linear combinations of the  $d_{xz}$, $d_{yz}$, $d_{xy}$, and $d_{x^2 -y^2}$ orbitals from the platinum. In particular, they will dominate around the K and M points, and their contributions can be related to the existence of non-vanishing $\sigma_{OH}^{L_x}$ and $\sigma_{OH}^{L_y}$ responses as reported in table \ref{table_1TTMD_Sup}.

\section{Relation between Pseudo-time-reversal and the $L_z$ operator for 2H TMDs}

Before showing the relationship between the operator for the calculation of the orbital Chern number and the pseudo-time-reversal operator, let us write the latter. Since the $d_Z$ orbitals belong to a $1\times1$ irreducible representation ($A^\prime_1$), they will transform as a scalar under the pseudo-time-reversal symmetry. However, following the procedure used in reference \cite{mei2016pseudo}, we can use the rotation operators to determine the pseudo-time-reversal operations of the states $d_{xy}$ and $d_{x^2-y^2}$, which belong to the irreducible representation $E^\prime$ of the group $D_{3h}$. In the $\left\lbrace |d_{z^2}\rangle,|d_{xy}\rangle, |d_{x^2 - y^2}\rangle \right \rbrace$ basis, the rotation operators for the 1H TMDs reads:

\begin{eqnarray}
\label{R_z}
R_z\left(\frac{2\pi}{3}\right) = \begin{pmatrix}
    1 & 0 & 0 \\
    0 & -\frac{1}{2} & \frac{\sqrt{3}}{2}\\
    0 & -\frac{\sqrt{3}}{2} & -\frac{1}{2}
\end{pmatrix} \qquad &R_z\left(-\frac{2\pi}{3}\right) = \begin{pmatrix}
    1 & 0 & 0 \\
    0 & -\frac{1}{2} & -\frac{\sqrt{3}}{2}\\
    0 & \frac{\sqrt{3}}{2} & -\frac{1}{2}
\end{pmatrix} \qquad 
\end{eqnarray}
where the operator $R_z (\phi)$ represents a rotation operator of angle $\phi$ around the $\hat{z}$-axis. The  rotation operators are block-diagonal, evincing that the $d_{z^2}$, and the orbitals $d_{xy}$ and $d_{x^2-y^2}$ belong to different irreducible representations. Moreover, the difference between the two rotation operators defined in Eq. \ref{R_z} is restricted to the $2\times2$ subspace spanned by the orbitals $d_{xy}$ and $d_{x^2-y^2}$, which may combine, to form a pair of pseudo-spins. Therefore for this subspace, we may define the pseudo-time-reversal operator $\mathcal{T}_p$=$U\mathcal{K}$, where the operator $\mathcal{K}$ represents the complex conjugation operator, and $U$ represents the unitary part of the pseudo-time-reversal symmetry operator  given by:

\begin{equation}\label{eqn:Uptr}
    U = \frac{1}{\sqrt{3}}\left(R_z\left(\frac{2\pi}{3}\right) -R_z\left(\frac{-2\pi}{3}\right)\right) = i\sigma_y
\end{equation}
where $\sigma_y$ is the usual Pauli matrix acting in the pseudo-spins' subspace.

Now that we have defined the pseudo-time-reversal operator, one can verify  that when applied to valence band states $|\psi_\tau\rangle = \big( \big|d^1_{x^2-y^2}\rangle-i\tau \big|d^1_{xy}\rangle \big)/\sqrt{2}$ at the $K$ point, it translates them to the other valley, similarly to the time-reversal operator. As far as the connection between the pseudo-time-reversal and orbital angular momentum operator is concerned, we may expand $U$ given by Eq. \ref{eqn:Uptr} in powers of $L_z$  to show that it is proportional to $L_z$. Hence, one can redefine the operator used in the orbital Chern number calculation presented in reference \cite{Us3} as the difference between the rotation operator and its time-reversed partner. 

\section{Effect of \lowercase{\textit{h}}-BN substrate}

\begin{figure}[!h]
\begin{center}
\includegraphics[width=0.65\textwidth]{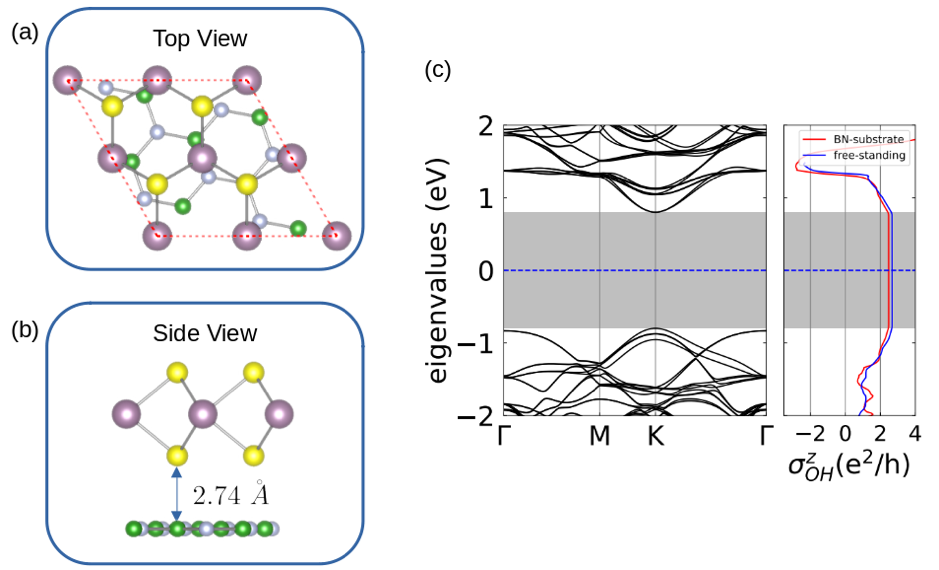}
\end{center}
\caption{Top view (a) and side view (b) of  MoS$_2$ on hBN substrate. (c) Band structure and orbital Hall conductivity of the MoS$_2$-\textit{h}BN bilayer.\label{Fig-substrate} }
\end{figure}

To study the effect of substrate usually present in typical experiments on orbital transport, we performed DFT calculations for a single layer of H-MoS$_2$ on top of a hexagonal boron nitride layer (hBN). Due to the lattice mismatch, $a_{\rm{MoS2}}=$ 3.18 \AA{} and $a_{\rm{hBN}}$ = 2.51 \AA, we constructed a supercell with a MoS$_2$ 2x2  and an hBN $\sqrt{7} \times \sqrt{7}$ to reduce the  mismatch. As a result, our lattice mismatch is smaller than 5$\%$. We preserved the MoS$_2$ structure and strained the hBN substrate (we also did the opposite and obtained similar results, with a small change in the band gap). The top and side views of the structure are shown in Fig. \ref{Fig-substrate}, panels (a) and (b), respectively. To correctly describe the dispersion forces between the two materials we included a Van Der Waals correction, resulting in a 2.74 \AA{} interlayer distance. In Fig. \ref{Fig-substrate} (c) we show its band structure along with the OHC (red line). The OHC for the free-standing H-MoS$_2$ monolayer (represented by the blue line) is included for comparison. We note that the OHC inside the insulating gap is practically  unaffected by the substrate.

\end{document}